\documentclass[preprint, superscriptaddress, draft, amssymb, floats, byrevetex]{revtex4}
\usepackage{latexsym}
\usepackage{graphics}

\newcommand{\bmat}{\left(\begin{array}}
\newcommand{\emat}{\end{array}\right)}
\newcommand{\beq}{\begin{equation}}
\newcommand{\eeq}{\end{equation}}

\newcommand{\drawsquare}[2]{\hbox{%
\rule{#2pt}{#1pt}\hskip-#2pt
\rule{#1pt}{#2pt}\hskip-#1pt
\rule[#1pt]{#1pt}{#2pt}}\rule[#1pt]{#2pt}{#2pt}\hskip-#2pt
\rule{#2pt}{#1pt}}

\newcommand{\fund}{\raisebox{-.5pt}{\drawsquare{6.5}{0.4}}}
\newcommand{\Ysymm}{\raisebox{-.5pt}{\drawsquare{6.5}{0.4}}\hskip-0.4pt%
        \raisebox{-.5pt}{\drawsquare{6.5}{0.4}}}
\newcommand{\Yasymm}{\raisebox{-3.5pt}{\drawsquare{6.5}{0.4}}\hskip-6.9pt%
        \raisebox{3pt}{\drawsquare{6.5}{0.4}}}
\newcommand{\antifund}{\overline{\fund}}

\def\yzero{\smash{\hbox{$y\kern-4pt\raise1pt\hbox{${}^\circ$}$}}}

\def\-{\hphantom{-}}
\def\ov{\overline}
\def\s2{\frac{1}{\sqrt2}}

\def\beq{\begin{equation}}
\def\eeq{\end{equation}}
\def\beqa{\begin{eqnarray}}
\def\eeqa{\end{eqnarray}}

\def\IF{\relax{\rm I\kern-.18em F}}
\def\II{\relax{\rm I\kern-.18em I}}
\def\IP{\relax{\rm I\kern-.18em P}}

\def\Dsl{\,\raise.15ex\hbox{/}\mkern-13.5mu D} 

\def\IC{\bf C}
\def\IZ{\bf Z}
\def\IT{\bf T}
\def\z2z2{$\IC^3/(\IZ_2\times\IZ_2)$}

\def\s{\sigma}

\def\z{\zeta}

\def\bo{{\raise-.3ex\hbox{\large$\Box$}}}               
\def\face{{\raise.2ex\hbox{$\displaystyle \bigodot$}\mskip-2.2mu \llap {$\ddot
        \smile$}}}                                      

\def\leftrightarrowfill{$\mathsurround=0pt \mathord\leftarrow \mkern-6mu
        \cleaders\hbox{$\mkern-2mu \mathord- \mkern-2mu$}\hfill
        \mkern-6mu \mathord\rightarrow$}       
\def\dvec#1{\vbox{\ialign{##\crcr
        \leftrightarrowfill\crcr\noalign{\kern-1pt\nointerlineskip}
        $\hfil\displaystyle{#1}\hfil$\crcr}}}           

\def\beq{\begin{equation}}
\def\eeq{\end{equation}}

\def\beqx{\begin{displaymath}}
\def\eeqx{\end{displaymath}}

\def\beqa{\begin{eqnarray}}
\def\eeqa{\end{eqnarray}}

\begin{document}

\title{
\normalsize \mbox{ }\hspace{\fill}
\begin{minipage}{12 cm}
{\tt ~~~~~~~~~~~~~~~~~~
UPR-1068-T,
 hep-th/0403061}{\hfill}
\end{minipage}\\[5ex]
{\large\bf   Supersymmetric Pati-Salam Models from Intersecting
D6-branes: A Road to the Standard Model
\\[1ex]}}
\date{\today}
\author{Mirjam Cveti\v c}
\affiliation{ Department of Physics and Astronomy, University of Pennsylvania, \\
Philadelphia, PA 19104, USA}
\author{Tianjun Li}
\affiliation{School of Natural Science, Institute for Advanced Study,  \\
             Einstein Drive, Princeton, NJ 08540, USA}
\author{Tao Liu}
\affiliation{ Department of Physics and Astronomy, University of Pennsylvania, \\
Philadelphia, PA 19104, USA}



\begin{abstract}

We provide a systematic construction of  three-family $N=1$
supersymmetric  Pati-Salam models from Type IIA orientifolds on
$\IT^6/(\IZ_2\times \IZ_2)$ with intersecting D6-branes. All the
gauge symmetry factors $SU(4)_C\times SU(2)_L \times SU(2)_R$
arise from the stacks of D6-branes with $U(n)$  gauge symmetries,
while the ``hidden sector'' is specified by $USp(n)$ branes,
parallel with the orientifold planes or their ${\bf Z_2}$ images.
The Pati-Salam gauge symmetry can be broken down to the
$SU(3)_C\times SU(2)_L\times U(1)_{B-L} \times U(1)_{I_{3R}}$ via
D6-brane splittings, and further down to the Standard Model  via
D- and F-flatness preserving Higgs mechanism from massless open
string states in a $N=2$ subsector. The models also possess  at
least two confining hidden gauge sectors,   where  gaugino
condensation can in turn trigger supersymmetry breaking and (some)
moduli stabilization. The systematic search  yields  11
inequivalent models: 8 models with less than 9 Standard model
Higgs doublet-pairs  and  1 model with only 2 Standard Model Higgs
doublet-pairs, 2 models possess at the string scale the gauge coupling
unification of $SU(2)_L$ and $SU(2)_R$, and all the models possess
additional exotic matters. We also make preliminary comments on
phenomenological implications of these models.

\end{abstract} \maketitle

\newpage

\section{Introduction}
Prior to the second string revolution the efforts in string
phenomenology focused  on  constructions of four-dimensional
solutions  in the weakly coupled heterotic string theory: the goal
was to construct $N=1$ supersymmetric models with features of the
Standard Model (SM).
On the other hand, the M-theory unification possesses in addition
to its  perturbative heterotic string theory corner, also other
corners such as perturbative Type I, Type IIA and Type IIB
superstring theory, which should provide new  potentially
phenomenologically interesting four-dimensional string solutions,
related to the heterotic ones via a web of  string dualities.  In
particular, the advent of D-branes \cite{JPEW}, as boundaries of
open strings,  plays an important  role in constructions of
phenomenologically interesting models in  Type I, Type IIA and
Type IIB string theories.
Conformal field theory techniques in the open string sectors,
which end on D-branes, allow for exact constructions of consistent
4-dimensional supersymmetric $N=1$ chiral models with non-Abelian
gauge symmetry on Type II orientifolds. Within this framework
chiral matters  can appear (i) due to D-branes located at orbifold
singularities with chiral fermions appearing on the worldvolume of
such D-branes [2$-$8] and/or (ii) at the intersections of D-branes
in the internal space~\cite{bdl} (These latter models also have a
T-dual description in terms of magnetized
D-branes~\cite{bachas,urangac}.).

Within the models with  intersecting D6-brane on Type IIA
orientifolds [12$-$14], a large number of
 non-supersymmetric three-family
Standard-like models and grand unified models were constructed
[12$-$25]. These models satisfy the Ramond-Ramond (RR) tadpole
cancellation conditions, however, since the models are
non-supersymmetric, there are uncancelled
Neveu-Schwarz-Neveu-Schwarz (NS-NS) tadpoles. In addition, the
string scale is close to the Planck scale because the intersecting
D6-branes typically have no common transverse direction in the
internal space. Therefore, these models typically suffer from the
large Planck scale corrections at the loop level, {\it i.e.},
there exists the gauge hierarchy problem.

On the other hand, the supersymmetric models~\cite{CSU1,CSU2} with
quasi-realistic features of the supersymmetric Standard-like
models have been constructed in Type IIA theory on
$T^6/(\IZ_2\times \IZ_2)$ orientifold with intersecting D6-branes.
Subsequently, a larger set of supersymmetric Standard-like models
and a Pati-Salam model~\cite{CP}, as well as a systematic
construction of supersymmetric $SU(5)$ Grand Unified models
\cite{CPS} have been constructed. Their  phenomenological
consequences, such as renormalization group running for the gauge
couplings, supersymmetry breaking via gaugino condensations,
moduli stabilization, and the complete Yukawa couplings that
include classical and quantum contributions, have been
studied~\cite{CLS1,CLS2,MCIP,CLW}. Furthermore, the supersymmetric
Pati-Salam models based on $Z_4$ and $Z_4\times Z_2$ orientifolds
with intersecting D6-branes were also
constructed~\cite{blumrecent,Honecker}. In these models, the
left-right gauge symmetry was obtained via brane recombinations,
so the final models do not have an explicit toroidal orientifold
construction, where the conformal field theory can be applied for
the calculation of the full spectrum and couplings.

We shall concentrate on constructions of supersymmetric
three-family Pati-Salam models based on   $T^6/(\IZ_2\times
\IZ_2)$ orientifold. Although previous constructions provided a
number of supersymmetric three family examples with Standard-like
gauge group (or Grand Unified group), these models have a number
of phenomenological problems. Previous models had at least 8 pairs
of the SM Higgs doublets and a number of exotic particles, some of
them fractionally charged. In addition, in the previous
supersymmetric SM constructions~\cite{CSU1,CSU2,CP}, there are at
least two extra anomaly free $U(1)$ gauge symmetries. They could
be in principle spontaneously broken via Higgs mechanism by the
scalar components of the chiral superfields with  the quantum
numbers of the right-handed neutrinos, however they  break
D-flatness conditions, and thus supersymmetry, so the scale of
symmetry breaking should be near the  electroweak scale. On the
other hand, there were typically no candidates,  preserving
D-flatness and F-flatness conditions, which could break these
gauge symmetries at an intermediate scale. In addition, there is
no gauge coupling unification. (The gauge coupling unification can
be realized in the quasi-supersymmetric $U(n)^3$
models~\cite{LLG3}. But the filler branes, which are on top of
orientifold planes in these models, are anti-branes and break the
supersymmetry at the string scale.) Furthermore, there exist three
multiplets in the adjoint representation for each $U(n)$ group,
which is a generic property for the supersymmetric and
non-supersymmetric toroidal orientifold constructions because the
typical three cycles wrapped by D6-branes are not rigid. One
possible solution is that we consider the Calabi-Yau
compactifications with rigid supersymmetric cycles, but, the
calculational techniques of conformal field theory may not be
applicable there. (For studies of $N=1$ supersymmetric  solutions
of D-branes on Calabi-Yau manifolds see \cite{RB11,RB12,Ilke}  and
references therein.)

Well-motivated by the Standard Model constructions,
 we shall study systematically  the three family $N=1$ supersymmetric
Pati-Salam models from Type IIA orientifolds on
$\IT^6/(\IZ_2\times \IZ_2)$ with intersecting D6-branes where all
the gauge symmetries come from $U(n)$ branes. On the one hand,
Pati-Salam model provides a natural origin of $U(1)_{B-L}$ and
$U(1)_{I_{3R}}$ both of which are generically required due to the
quantum numbers of the SM fermions and the hypercharge interaction
in the SM building from intersecting D6-brane scenarios. On the
other hand, it also provides one road to the SM without any
additional  anomaly-free $U(1)$'s near the electroweak scale,
which was a generic feature  of  the previous supersymmetric SM
constructions~\cite{CP}.

The paper is organized in the following way. In Section II  we
briefly review the rules for supersymmetric model building with
intersecting D6-branes on Type IIA orientifolds,  the conditions
for the tadpole cancellations and conditions on D6-brane
configurations for $N=1$ supersymmetry in four-dimension. We
specifically focus on Type IIA theory on $T^6/(\IZ_2\times \IZ_2)$
orientifold.

In Section III, we discuss in detail the T-duality symmetry and
its variations within the supersymmetric intersecting D6-brane
model building.  The first set of symmetries  are general and can
be applied to any concrete particle physics model building (type
I), and the second set is special and only valid for the specific
Pati-Salam model building  (type II). We also find that any two
models T-dual to each other have the same gauge couplings at
string scale.

In Section IV, we study the phenomenological constraints in the
construction of the supersymmetric Pati-Salam models. In
particular, we highlight why the models where the Pati-Salam gauge
symmetry comes from $U(n)$ branes are phenomenologically
interesting. We show that the Pati-Salam gauge symmetry can be
broken down to the $SU(3)_C\times SU(2)_L\times U(1)_{B-L} \times
U(1)_{I_{3R}}$ via D6-brane splittings, and further down to the
Standard Model gauge symmetry via Higgs mechanism, $i.e.$, brane
recombination in geometric interpretation~\cite{CIM2}, where the
Higgs fields come from the massless open string states in a $N=2$
subsector. In order to stabilize the moduli and provide a way to
break supersymmetry, we require that at least two hidden sector
gauge factors are confining, thus allowing for the ``race-track''
gaugino condensation mechanism.

In Section V, by employing the T-duality and its variations, we
systematic search for the inequivalent models, and  discuss the
possible classes of solutions in detail. As a result, we obtain
total of 11 inequivalent models,  all of them arising from the
orbifold with only one of the three two-tori tilted. Compared to
the previous SM constructions~\cite{CSU1,CSU2,CP}, eight of our
models have fewer pairs of the SM Higgs doublets ($\leq 8$).
 Interestingly, the gauge coupling unification for $SU(2)_L$ and
$SU(2)_R$ can be achieved at string scale in two models. As
explicit examples, we present the chiral spectra in the open
string sector for the models I-NZ-1a, I-Z-6 and I-Z-10.

In Section VI, we briefly comment on the other potentially
interesting setups. And the discussions and conclusions are given
in Section VII. In Appendix, we present the D6-brane
configurations and intersection numbers for supersymmetric
Pati-Salam models.

\section{Conditions  for Supersymmetric Models  from
$T^6 /(Z_2 \times Z_2)$ Orientifolds with Intersecting D6-Branes}

The rules to construct supersymmetric models from Type IIA
orientifolds on $T^6 /(Z_2 \times Z_2)$ with D6-branes
intersecting at generic angles, and to obtain the spectrum of
massless open string states have been discussed in~\cite{CSU2}.
Following the convention in Ref.~\cite{CPS}, we briefly review the
essential points  in the construction of such  models.

The starting point is Type IIA string theory compactified on a
$T^6 /(Z_2 \times Z_2)$ orientifold. We consider $T^{6}$ to be a
six-torus factorized as $T^{6} = T^{2} \times T^{2} \times T^{2}$
whose complex coordinates are $z_i$, $i=1,\; 2,\; 3$ for the
$i$-th two-torus, respectively. The $\theta$ and $\omega$
generators for the orbifold group $Z_{2} \times Z_{2}$, which are
associated with their twist vectors $(1/2,-1/2,0)$ and
$(0,1/2,-1/2)$ respectively, act on the complex coordinates of
$T^6$ as \beqa
& \theta: & (z_1,z_2,z_3) \to (-z_1,-z_2,z_3)~,~ \nonumber \\
& \omega: & (z_1,z_2,z_3) \to (z_1,-z_2,-z_3)~.~\,
\label{orbifold} \eeqa The orientifold projection is implemented
by gauging the symmetry $\Omega R$, where $\Omega$ is world-sheet
parity, and $R$ acts as \beqa
 R: (z_1,z_2,z_3) \to ({\ov z}_1,{\ov z}_2,{\ov
z}_3)~.~\, \label{orientifold} \eeqa So, there are four kinds of
orientifold 6-planes (O6-planes) for the actions of $\Omega R$,
$\Omega R\theta$, $\Omega R \omega$, and $\Omega R\theta\omega$,
respectively. To cancel the RR charges of O6-planes, we introduce
stacks of $N_a$ D6-branes, which wrap on the factorized
three-cycles. Meanwhile, we have two kinds of complex structures
consistent with orientifold projection for a two-torus --
rectangular and tilted~\cite{bkl,CSU2,CPS}. If we denote the
homology classes of the three cycles wrapped by the D6-brane
stacks as $n_a^i[a_i]+m_a^i[b_i]$ and $n_a^i[a'_i]+m_a^i[b_i]$
with $[a_i']=[a_i]+\frac{1}{2}[b_i]$ for the rectangular and
tilted tori respectively, we can label a generic one cycle by
$(n_a^i,l_a^i)$ in either case, where in terms of the wrapping
numbers $l_{a}^{i}\equiv m_{a}^{i}$ for a rectangular two-torus
and $l_{a}^{i}\equiv 2\tilde{m}_{a}^{i}=2m_{a}^{i}+n_{a}^{i}$ for
a tilted two-torus. Note that for a tilted two-torus,
$l_a^i-n_a^i$ must be even. For a stack of $N_a$ D6-branes along
the cycle $(n_a^i,l_a^i)$, we also need to include their $\Omega
R$ images $N_{a'}$ with wrapping numbers $(n_a^i,-l_a^i)$. For
D6-branes on top of O6-planes, we count the D6-branes and their
images independently. So, the homology three-cycles for stack $a$
of $N_a$ D6-branes and its orientifold image $a'$ take the form
\beq
[\Pi_a]=\prod_{i=1}^{3}\left(n_{a}^{i}[a_i]+2^{-\beta_i}l_{a}^{i}[b_i]\right),\;\;\;
\left[\Pi_{a'}\right]=\prod_{i=1}^{3}
\left(n_{a}^{i}[a_i]-2^{-\beta_i}l_{a}^{i}[b_i]\right)~,~\, \eeq
where $\beta_i=0$ if the $i$-th two-torus is rectangular and
$\beta_i=1$ if it is tilted. And the homology three-cycles wrapped
by the four O6-planes are \beq \Omega R: [\Pi_{\Omega R}]= 2^3
[a_1]\times[a_2]\times[a_3]~,~\, \eeq \beq \Omega R\omega:
[\Pi_{\Omega
R\omega}]=-2^{3-\beta_2-\beta_3}[a_1]\times[b_2]\times[b_3]~,~\,
\eeq \beq \Omega R\theta\omega: [\Pi_{\Omega
R\theta\omega}]=-2^{3-\beta_1-\beta_3}[b_1]\times[a_2]\times[b_3]~,~\,
\eeq \beq
 \Omega R\theta:  [\Pi_{\Omega
R}]=-2^{3-\beta_1-\beta_2}[b_1]\times[b_2]\times[a_3]~.~\,
\label{orienticycles} \eeq Therefore, the intersection numbers are
\beq
I_{ab}=[\Pi_a][\Pi_b]=2^{-k}\prod_{i=1}^3(n_a^il_b^i-n_b^il_a^i)~,~\,
\eeq \beq
I_{ab'}=[\Pi_a]\left[\Pi_{b'}\right]=-2^{-k}\prod_{i=1}^3(n_{a}^il_b^i+n_b^il_a^i)~,~\,
\eeq \beq
I_{aa'}=[\Pi_a]\left[\Pi_{a'}\right]=-2^{3-k}\prod_{i=1}^3(n_a^il_a^i)~,~\,
\eeq \beq {I_{aO6}=[\Pi_a][\Pi_{O6}]=2^{3-k}(-l_a^1l_a^2l_a^3
+l_a^1n_a^2n_a^3+n_a^1l_a^2n_a^3+n_a^1n_a^2l_a^3)}~,~\,
\label{intersections} \eeq where $[\Pi_{O6}]=[\Pi_{\Omega
R}]+[\Pi_{\Omega R\omega}]+[\Pi_{\Omega
R\theta\omega}]+[\Pi_{\Omega R\theta}]$ is the sum of O6-plane
homology three-cycles wrapped by the four O6-planes,
 and $k=\beta_1+\beta_2+\beta_3$ is
the total number of tilted two-tori. For future convenience, we
shall define $n_{a}^il_b^i\pm n_b^il_a^i$ as the intersection
factor from the $i-$th two-torus.

\begin{table}[t]
\caption{General spectrum on intersecting D6-branes at generic
angles which is valid for both rectangular and tilted two-tori.
The representations in the table refer to $U(N_a/2)$, the
resulting gauge symmetry~\cite{CSU2} due to $Z_2\times Z_2$
orbifold projection. For supersymmetric constructions, scalars
combine with fermions to form chiral supermultiplets. In our
convention, positive intersection numbers implie left-hand chiral
supermultiplets.}
\renewcommand{\arraystretch}{1.25}
\begin{center}
\begin{tabular}{|c|c|}
\hline {\bf Sector} & \phantom{more space inside this box}{\bf
Representation}
\phantom{more space inside this box} \\
\hline\hline
$aa$   & $U(N_a/2)$ vector multiplet  \\
       & 3 adjoint chiral multiplets  \\
\hline
$ab+ba$   & $I_{ab}$ $(\fund_a,\antifund_b)$ fermions   \\
\hline
$ab'+b'a$ & $I_{ab'}$ $(\fund_a,\fund_b)$ fermions \\
\hline $aa'+a'a$ &$\frac 12 (I_{aa'} - \frac 12 I_{a,O6})\;\;
\Ysymm\;\;$ fermions \\
          & $\frac 12 (I_{aa'} + \frac 12 I_{a,O6}) \;\;
\Yasymm\;\;$ fermions \\
\hline
\end{tabular}
\end{center}
\label{spectrum}
\end{table}

The general spectrum of D6-branes intersecting at generic angles,
which is valid for both rectangular and tilted two-tori, is given
in Table \ref{spectrum}. And the 4-dimensional $N=1$
supersymmetric models from Type IIA orientifolds with intersecting
D6-branes are mainly constrained in two aspects: RR tadpole
cancellation (section 2.1) and $N=1$ supersymmetry in four
dimensions (section 2.2).

\subsection{Tadpole Cancellation Conditions}

As sources of RR fields, D6-branes and orientifold O6-planes are
required to satisfy the Gauss law in a compact space, {\it i.e.},
the total RR charges of D6-branes and O6-planes must vanish since
the RR field flux lines are conserved. The RR tadpole cancellation
conditions are
\begin{eqnarray}
\sum_a N_a [\Pi_a]+\sum_a N_a
\left[\Pi_{a'}\right]-4[\Pi_{O6}]=0~,~\,
\end{eqnarray}
where the last contribution comes from the O6-planes which have
$-4$ RR charges in the D6-brane charge unit.

To simplify the notation, let us define the products of wrapping
numbers \beq
\begin{array}{rrrr}
A_a \equiv -n_a^1n_a^2n_a^3, & B_a \equiv n_a^1l_a^2l_a^3,
& C_a \equiv l_a^1n_a^2l_a^3, & D_a \equiv l_a^1l_a^2n_a^3, \\
\tilde{A}_a \equiv -l_a^1l_a^2l_a^3, & \tilde{B}_a \equiv
l_a^1n_a^2n_a^3, & \tilde{C}_a \equiv n_a^1l_a^2n_a^3, &
\tilde{D}_a \equiv n_a^1n_a^2l_a^3.\,
\end{array}
\label{variables}\eeq To cancel the RR tadpoles, we can also
introduce an arbitrary number of D6-branes wrapping cycles along
the orientifold planes, the so called ``filler branes'', which
contribute to the tadpole conditions but trivially satisfy the
supersymmetry conditions. Thus, the tadpole conditions are
\begin{eqnarray}
 -2^k N^{(1)}+\sum_a N_a A_a=-2^k N^{(2)}+\sum_a N_a
B_a= \nonumber\\ -2^k N^{(3)}+\sum_a N_a C_a=-2^k N^{(4)}+\sum_a
N_a D_a=-16,\,
\end{eqnarray}
where $2 N^{(i)}$ are the number of filler branes wrapping along
the $i$-th O6-plane which is defined in Table \ref{orientifold}.

The tadpole cancellation conditions directly lead to the
$SU(N_a)^3$ cubic non-Abelian anomaly
cancellation~\cite{Uranga,imr,CSU2}. And the cancellation of U(1)
mixed gauge and gravitational anomaly or $[SU(N_a)]^2 U(1)$ gauge
anomaly can be achieved by Green-Schwarz mechanism mediated by
untwisted RR fields~\cite{Uranga,imr,CSU2}.

\renewcommand{\arraystretch}{1.4}
\begin{table}[t]
\caption{Wrapping numbers of the four O6-planes.} \vspace{0.4cm}
\begin{center}
\begin{tabular}{|c|c|c|}
\hline
  Orientifold Action & O6-Plane & $(n^1,l^1)\times (n^2,l^2)\times
(n^3,l^3)$\\
\hline
    $\Omega R$& 1 & $(2^{\beta_1},0)\times (2^{\beta_2},0)\times
(2^{\beta_3},0)$ \\
\hline
    $\Omega R\omega$& 2& $(2^{\beta_1},0)\times (0,-2^{\beta_2})\times
(0,2^{\beta_3})$ \\
\hline
    $\Omega R\theta\omega$& 3 & $(0,-2^{\beta_1})\times
(2^{\beta_2},0)\times
(0,2^{\beta_3})$ \\
\hline
    $\Omega R\theta$& 4 & $(0,-2^{\beta_1})\times (0,2^{\beta_2})\times
    (2^{\beta_3},0)$ \\
\hline
\end{tabular}
\end{center}
\label{orientifold}
\end{table}

\subsection{Conditions for 4-Dimensional $N = 1$ Supersymmetric D6-Brane}

The four-dimensional $N=1$ supersymmetric models require that 1/4
supercharges from ten-dimensional Type I T-dual be preserved, {\it
i.e.}, these $1/4$ supercharges should survive two
 projections: the orientation projection of the
intersecting D6-branes, and the $Z_2\times Z_2$ orbifold
projection on the background manifold. Analysis shows that, the
4-dimensional $N=1$ supersymmetry (SUSY) can be preserved by the
orientation projection if and only if the rotation angle of any
D6-brane with respect to the orientifold-plane is an element of
$SU(3)$\cite{bdl}, or in other words,
$\theta_1+\theta_2+\theta_3=0 $ mod $2\pi$, where $\theta_i$ is
the angle between the $D6$-brane and the orientifold-plane in the
$i$-th two-torus. Meanwhile, this 4-dimensional $N=1$
supersymmetry will automatically survive the $Z_2\times Z_2$
orbifold projection. The SUSY conditions can therefore be written
as~\cite{CPS}
\begin{eqnarray}
x_A\tilde{A}_a+x_B\tilde{B}_a+x_C\tilde{C}_a+x_D\tilde{D}_a=0,
\nonumber\\\nonumber \\ A_a/x_A+B_a/x_B+C_a/x_C+D_a/x_D<0,
\label{susyconditions}
\end{eqnarray} where $x_A=\lambda,\;
x_B=\lambda 2^{\beta_2+\beta3}/\chi_2\chi_3,\; x_C=\lambda
2^{\beta_1+\beta3}/\chi_1\chi_3,\; x_D=\lambda
2^{\beta_1+\beta2}/\chi_1\chi_2$, and $\chi_i=R^2_i/R^1_i$ are the
complex structure moduli. The positive parameter $\lambda$ has
been introduced to put all the variables $A,\,B,\,C,\,D$ on an
equal footing. Based on these SUSY conditions,  all possible
D6-brane configurations preserving 4-dimensional $N=1$
supersymmetry can be classified into three types:

(1) Filler brane with the same wrapping numbers as one of the
O6-planes in Table \ref{orientifold}. It corresponds to the $USp$
group. And among coefficients $A$, $B$, $C$ and $D$, one and only
one of them is non-zero and negative. If the filler brane has
non-zero $A$, $B$, $C$ or $D$, we refer to the $USp$ group as the
$A$-, $B$-, $C$- or $D$-type $USp$ group, respectively.

(2) Z-type D6-brane which contains one zero wrapping number. Among
$A$, $B$, $C$ and $D$, two are negative and two are zero.

(3) NZ-type D6-brane which contains no zero wrapping number. Among
$A$, $B$, $C$ and $D$, three are negative and the other one is
positive. Based on which one is positive, these NZ-type branes are
defined as the $A$-, $B$-, $C$- and $D$-type NZ brane. Each type
can have two forms of the wrapping numbers which are defined as
\begin{eqnarray}
A1: (-,-)\times(+,+)\times(+,+),& A2:(-,+)\times(-,+)\times(-,+);\\
B1: (+,-)\times(+,+)\times(+,+),& B2:(+,+)\times(-,+)\times(-,+);\\
C1: (+,+)\times(+,-)\times(+,+),& C2:(-,+)\times(+,+)\times(-,+);\\
D1: (+,+)\times(+,+)\times(+,-),& D2:(-,+)\times(-,+)\times(+,+).
\end{eqnarray}
In the following, we'll call the Z-type and NZ-type D6-branes as
 $U$-branes since they carry $U(n)$ gauge symmetry.

\section{T-Duality Symmetry and its Variations}

T-duality relates equivalent models, and thus employing this
symmetry can simplify  significantly  the search for the
inequivalent models. In this section, we shall study  the action
of the T-dualities (and its variants). These symmetries correspond
to  two types: one is general and can be applied to any D6-brane
model building (type I) and the other one is special and only
effective in the Pati-Salam model building (type II). Our
philosophy is: consider only one model for each equivalent class
characterized by these T-dualities.

Before discussing the T-duality, we point out that: (1) Two models
are equivalent if their three two-tori and the corresponding
wrapping numbers for all the D6-branes are related by an element
of permutation group $S_3$ which acts on three two-tori. This is a
trivial fact. (2) Two D6-brane configurations are equivalent if
their wrapping numbers on two arbitrary two-tori have the same
magnitude but opposite sign, and their wrapping numbers on the
third two-torus are the same. We call it as the D6-brane Sign
Equivalent Principle (DSEP). In the following, we discuss type I
and type II T-dualities separately:\\

Type I T-duality: T-duality transformation happens on two two-tori
simultaneously, for example, the $j$-th and $k$-th two-tori
\begin{eqnarray}
(n_x^j, l_x^j) \longrightarrow (-l_x^j, n_x^j), & (n_x^k, l_x^k)
\longrightarrow (l_x^k, - n_x^k)~,~\, \label{T-duality I}
\end{eqnarray}
where $x$ runs over all stacks of D6-branes in the model. It
doesn't change anything about the D-brane model except that it
makes an interchange among the four pairs of the products of
wrapping numbers $(A,\tilde{A})$, $(B,\tilde{B})$, $(C,\tilde{C})$
and $(D,\tilde{D})$, which indicates that the particle spectra are
invariant under this transformation while the complex structure
moduli may not be.

Without loss of generality, we assume that $j=2$ and $k=3$. In
this case, the interchange will take place between A and B pairs,
and C and D pairs
\begin{eqnarray}
(A,\tilde{A})\leftrightarrow(B,\tilde{B}),&
(C,\tilde{C})\leftrightarrow(D,\tilde{D}).\,
\end{eqnarray}
And the corresponding transformations of moduli parameters are
\begin{eqnarray}
x_A'=x_B, ~x_B'=x_A, ~x_C'=x_D, ~x_D'=x_C,\,
\end{eqnarray}
After these transformations, we obtain the new complex structure
moduli and radii of $T^6$
\begin{eqnarray}
\chi_1'=2^{\beta_1}\sqrt{{x_Ax_B}\over{x_Cx_D}}=\chi_1,\\
\chi_2'=2^{\beta_2}\sqrt{{x_Dx_B}\over{x_Ax_C}}=2^{2\beta_2}(\chi_2)^{-1},\\
\chi_3'=2^{\beta_3}\sqrt{{x_Cx_B}\over{x_Ax_D}}=2^{2\beta_3}(\chi_3)^{-1},
\end{eqnarray}
\begin{eqnarray}
(R_1^1)'=R_1^1,~(R_1^2)'=R_1^2,\\
(R_2^1)'={{2^{-\beta_2}M_s^2}\over{R_2^1}},~(R_2^2)'={{2^{\beta_2}M_s^2}\over{R_2^2}},\\
(R_3^1)'={{2^{-\beta_3}M_s^2}\over{R_3^1}},~(R_3^2)'={{2^{\beta_3}M_s^2}\over{R_3^2}},
\end{eqnarray}
where $M_s$ is the string scale.

Sometimes, it will be more convenient if we combine this T-duality
with the trivial two two-tori interchange $T_j \leftrightarrow
T_k$ where the transformations of the wrapping numbers and
$\beta_{j, k}$ are
\begin{eqnarray}
n_x^j \leftrightarrow n_x^k,& l_x^j \leftrightarrow l_x^k,\\
\beta_j\leftrightarrow \beta_k.\,
\end{eqnarray}
Thus, under this extended T-duality, the wrapping numbers and
$\beta_{j, k}$ transform as
\begin{eqnarray}
(n_x^j, l_x^j) \longrightarrow (l_x^k, - n_x^k), &
(n_x^k, l_x^k) \longrightarrow (-l_x^j, n_x^j),\\
\beta_j\leftrightarrow \beta_k.\, \label{T-duality Ia}
\end{eqnarray}
Still for $j=2$ and $k=3$, only A and B pairs are interchanged
under this T-duality
\begin{eqnarray}
(A,\tilde{A})\leftrightarrow(B,\tilde{B}),~\,
\end{eqnarray}
and the corresponding transformations of moduli parameters are
\begin{eqnarray}
x_A'=x_B, x_B'=x_A, x_C'=x_C, x_D'=x_D.~\,
\end{eqnarray}
In this case, the new complex structure moduli and radii are
\begin{eqnarray}
\chi_1'=2^{\beta_1}\sqrt{{x_Ax_B}\over{x_Cx_D}}=\chi_1,~\,\\
\chi_2'=2^{\beta_3}\sqrt{{x_Bx_C}\over{x_Ax_D}}=2^{2\beta_3}\chi_3^{-1},~\,\\
\chi_3'=2^{\beta_2}\sqrt{{x_Bx_D}\over{x_Ax_C}}=2^{2\beta_2}\chi_2^{-1},~\,
\end{eqnarray}
\begin{eqnarray}
(R_1^1)'=R_1^1,~(R_1^2)'=R_1^2,~\,\\
(R_2^1)'={{2^{-\beta_3}M_s^2}\over{R_3^1}},~(R_2^2)'={{2^{\beta_3}M_s^2}\over{R_3^2}},~\,\\
(R_3^1)'={{2^{-\beta_2}M_s^2}\over{R_2^1}},~(R_3^2)'={{2^{\beta_2}M_s^2}\over{R_2^2}}.~\,
\end{eqnarray}
One pair of the models related by this extended type I T-duality
have been shown in Table \ref{model I-NZ-1a} and Table \ref{model
I-NZ-1b} in the Appendix up to DSEP on the first two two-tori of
$b$ stack of D6-branes. Since this extended T-duality only
interchanges two pairs of the products of wrapping numbers, if all
two-tori are rectangular or tilted, all models characterized by
the permutations of these four parameter pairs will be T-dual to
each other. By the way, for the case where two two-tori are
rectangular or tilted, this conclusion is not valid if the
rectangular and tilted two-tori have been fixed.

As a remark, in this kind of D6-brane models, the Yang-Mills gauge
coupling $g_{YM}^x$ for $x$-stack of D6-branes at string scale
is~\cite{CLW,LLG3}
\begin{eqnarray}
(g_{YM}^x)^2 = {{{\sqrt {8\pi}} M_s}\over\displaystyle {M_{Pl}}}
{1\over\displaystyle {\prod_{i=1}^3 {\sqrt {\left(n_x^{i}
\right)^2\chi_i^{-1} + \left(2^{-\beta_i} l_x^i \right)^2
\chi_i}}}}~,~\, \label{coupling}
\end{eqnarray}
where $M_{Pl}$ is the 4-dimensional Planck scale. So the gauge
coupling is invariant under the T-duality and its variation. This
is a typical property of T-duality(e.g. see \cite{Sen}).\\

Type II T-duality: Under it, the transformations of the wrapping
numbers  for any stacks of D6-branes in the model are
\begin{eqnarray}
n_x^i \rightarrow -n_x^i,~ l_x^i \rightarrow l_x^i,&
 n_x^j \leftrightarrow l_x^j,&
n_x^k \leftrightarrow l_x^k,\, \label{T-duality II}
\end{eqnarray}
where $i\not= j \not= k$, and
 $x$ runs over all D6-branes in the model.
Comparing with the general type I T-duality, besides the
interchanges among $(A,\tilde{A})$, $(B,\tilde{B})$,
$(C,\tilde{C})$ and $(D,\tilde{D})$, the signs of $\tilde{A}$,
$\tilde{B}$, $\tilde{C}$ and $\tilde{D}$ are also changed, which
lead to the sign changes of all intersection numbers. On the other
hand, for the models in our model construction, we require that
\begin{eqnarray}
I_{ab}+I_{ab'}=3, & I_{ac}=-3, & I_{ac'}=0,~ \,
\end{eqnarray}
which will be discussed in detail in the next Section. Obviously,
there is only one sign difference between the intersection numbers
of $I_{ab}$ and $I_{ac}$ if $I_{ab}=3$. By combining with
\begin{eqnarray}
b \leftrightarrow c,~\,
\end{eqnarray}
therefore, we may get one equivalent model satisfying our
requirements. For this type of T-duality, the moduli and radii
will obey the same transformation rules as those in the type I
T-duality. But, unlike the type I, the quantum numbers for
$SU(2)_L$ and $SU(2)_R$ in the particle spectrum and two gauge
couplings at string scale will be interchanged due to $b
\leftrightarrow c$. Models in Table \ref{model I-NZ-1a} and Table
\ref{model I-NZ-1c} are such a pair of examples related by type II
T-duality up to DSEP on $b$ and $c$ stacks of D6-branes.

If $I_{ab}=1$ or 2, which can be achieved only when $n_a^il_b^i=0$
or $n_b^il_a^i=0$ are satisfied on two two-tori, these
intersection numbers become
\begin{eqnarray}
I_{ab'}=3, ~I_{ac'}=-1 ~{\rm or}~ -2,\,
\end{eqnarray}
under the type II T-duality transformation. Therefore, if we relax
the intersection number requirement to
\begin{eqnarray}
I_{ab}+I_{ab'}=3,~ I_{ac}=-3,~I_{ac'}=0, \,
\end{eqnarray}
and
\begin{eqnarray}
I_{ab}=3,~I_{ab'}=0,~ I_{ac}+I_{ac'}=-3,\,
\end{eqnarray}
we only have to consider one case if the D6-brane wrapping numbers
in two setups can be related by Eq. (\ref{T-duality II}): the
derivations and conclusions in the first setup can be applied to
the second one as well.

By combining with type I T-duality and DSEP, we obtain an
variation of type II T-duality. Under it, the transformations of
the wrapping numbers for any stacks of D6-branes in the model are
\begin{eqnarray}
l_x^1 \rightarrow -l_x^1,& l_x^2 \rightarrow -l_x^2,& l_x^3
\rightarrow -l_x^3, \nonumber\\
b\leftrightarrow c, \label{T-duality IIa}
\end{eqnarray}
where $x$ runs over all D6-branes in the model. Since the
transformations in the first line of the above equations only
change the signs of $\tilde{A}$, $\tilde{B}$, $\tilde{C}$ and
$\tilde{D}$, the moduli and radii are invariant.

\section{Supersymmetric Pati-Salam Models and Gauge Symmetry
Breaking via D6-Brane Splittings}

To build the SM or SM-like models in the intersecting D6-brane
scenarios, besides $U(3)_C$ and $U(2)_L$ stacks of branes, we need
at least two extra $U(1)$ gauge symmetries for both SUSY and
non-SUSY versions due to the quantum number of the right-handed
electron~\cite{imr,CSU2,CPS,CP}. One ($U(1)_L$) is lepton number
symmetry, and the other one ($U(1)_{I_{3R}}$) is like the third
component of right-handed weak isospin. Then, the hypercharge is
obtained via
\begin{eqnarray}
Q_Y=Q_{I_{3R}}+{{Q_B-Q_{L}}\over{2}}~,~\,
\end{eqnarray}
where $U(1)_B$ arises from the overall $U(1)$ in $U(3)_C$.
Meanwhile, to forbid the gauge field of $U(1)_{I_{3R}}$ to obtain
a mass via $B\wedge F$ couplings, the $U(1)_{I_{3R}}$ can only
come from the non-Abelian part of $U(2)_R$ or $USp$ gauge
symmetry. In this case, the $U(1)$ gauge symmetry, which comes
from a non-Abelian symmetry, is generically anomaly free and its
gauge field is massless. Similarly, to generate the anomaly-free
$U(1)_{B-L}$ symmetry, the $U(1)_L$ should come from non-Abelian
group. Considering that the $U(1)_L$ stack should be parallel to
the $U(3)_C$ stack on at least one two-tori, we generate it by
splitting branes from one $U(4)$ stack, resulting in the $U(3)_C$
stack at the same time.

In the previous supersymmetric SM constructions~\cite{CSU2,CP},
$U(1)_{I_{3R}}$ arises from the stack of D6-branes on the top of
orientifold, {\it i.e.}, from the $USp$ group. These models have
at least 8 pairs of the SM Higgs doublets, and generically there
exist two additional anomaly free $U(1)$ gauge symmetries. They
could be in principle spontaneously broken via Higgs mechanism by
the scalar components of the chiral superfields with the quantum
numbers of the right-handed neutrinos, however they break
D-flatness conditions, and thus supersymmetry, so the scale of
symmetry breaking should be near the electroweak scale. On the
other hand, there were typically no candidates, preserving the
D-flatness and F-flatness conditions, and that could in turn break
these gauge symmetries at an intermediate scale.

Therefore, we focus on Pati-Salam model where $U(1)_{I_{3R}}$
arises from the $U(2)_R$ symmetry. Failing to find interesting
models with $SU(2)_L$ from the D6-branes on the top of O6-plane,
we would like to construct the supersymmetric $SU(4)_C\times
SU(2)_L\times SU(2)_R$ models from three stacks of D6-branes that
are not on the top of orientifold planes. We will show that the
Pati-Salam gauge symmetry can be broken down to $SU(3)_C\times
SU(2)_L\times U(1)_{B-L} \times U(1)_{I_{3R}}$ via D6-brane
splittings, and then down to the SM gauge symmetry via Higgs
mechanism where the Higgs particles come from a $N=2$ subsector.
In particular, in our models,
 we do not have any extra anomaly free U(1)
gauge symmetry at electroweak scale which was a generic problem in
previous constructions~\cite{CSU2,CP}.

Suppose we have three stacks of D6-branes, $a$, $b$, and $c$ with
number of D6-branes 8, 4, and 4. So, $a$, $b$, and $c$ stacks give
us the gauge symmetry $U(4)_C$, $U(2)_L$ and $U(2)_R$,
respectively. The anomalies from three $U(1)$s are cancelled by
the Green-Schwarz mechanism, and the gauge fields of these $U(1)$s
obtain masses via the linear $B\wedge F$ couplings. So, the
effective gauge symmetry is $SU(4)_C\times SU(2)_L\times SU(2)_R$.
In addition, we require that the intersection numbers satisfy
\begin{eqnarray}
\label{E3LF} I_{ab} + I_{ab'}~=~3~,~\,
\end{eqnarray}
\begin{equation}
\label{E3RF} I_{ac} ~=~-3~,~ I_{ac'} ~=~0~.~\,
\end{equation}
The conditions $I_{ab} + I_{ab'}=3$ and $I_{ac} =-3$ give us three
families of the SM fermions with quantum numbers $({\bf 4, 2, 1})$
and $({\bf {\bar 4}, 1, 2})$ under $SU(4)_C\times SU(2)_L\times
SU(2)_R$ gauge symmetry. $I_{ac'} =0 $ implies that $a$ stack of
D6-branes is parallel to the orientifold ($\Omega R$) image $c'$
of the $c$ stack of D6-branes along at least one tow-torus, for
example, the third two-torus. Then, there are open strings which
stretch between the $a$ and $c'$ stacks of D6-branes. If the
minimal distance squared $Z^2_{(ac')}$
 (in $1/M_s$ units) between these two stacks of D6-branes on the third
two-torus is small, {\it i.e.},  the minimal length squared of the
stretched string is small, we have the light scalars with
squared-masses $Z^2_{(ab')}/(4\pi^2 \alpha')$ from the NS sector,
and the light fermions with the same masses from the R
sector~\cite{Uranga,imr,LLG3}. These scalars and fermions form the
4-dimensional $N=2$ hypermultiplets, so, we obtain the
$I_{ac'}^{(2)}$ (the intersection numbers for $a$ and $c'$ stacks
on the first two two-tori) vector-pairs of the chiral multiplets
with quantum numbers $({\bf {\bar 4}, 1, 2})$ and $({\bf 4, 1,
2})$. These particles are the Higgs fields needed to break the
Pati-Salam gauge symmetry down to the SM gauge symmetry. In
particular, these particles are massless if $Z^2_{(ac')}=0$. By
the way, the intersection numbers $I_{ac}=0$ and $I_{ac'}=-3$ are
equivalent to $I_{ac}=-3$ and $I_{ac'}=0$ due to the symmetry
transformation $c\leftrightarrow c'$.

In order to break the gauge symmetry, we split the $a$ stack of
D6-branes into $a_1$ and $a_2$ stacks with 6 and 2 D6-branes,
respectively. The $U(4)_C$ gauge symmetry is broken down to the
$U(3) \times U(1)$. Let us assume that the numbers of symmetric
and anti-symmetric representations for $SU(4)_C$ are respectively
 $n_{\Ysymm}^a$ and $n_{\Yasymm}^a$, similar
convention for $SU(2)_L$ and $SU(2)_R$. After splitting, the gauge
fields and three multiplets in adjoint representation for
$SU(4)_C$ are broken down to  the gauge fields and three
multiplets in adjoint representations for $SU(3)_C\times
U(1)_{B-L}$, respectively. The $n_{\Ysymm}^a$ and $n_{\Yasymm}^a$
multiplets in symmetric and anti-symmetric representations for
$SU(4)_C$ are broken down to the $n_{\Ysymm}^a$ and
$n_{\Yasymm}^a$ multiplets in symmetric and anti-symmetric
representations for $SU(3)_C$, and $n_{\Ysymm}^a$ multiplets in
symmetric representation for $U(1)_{B-L}$. However, there are
$I_{a_1 a'_2}$ new fields with quantum number $({\bf 3, -1})$
under $SU(3)_C\times U(1)_{B-L}$ from the open strings at the
intersections of D6-brane stacks $a_1$ and $a_2'$. The rest of the
particle spectrum is the same. Moreover, we can show that the
anomaly free gauge symmetry from $a_1$ and $a_2$ stacks of
D6-branes is $SU(3)_C\times U(1)_{B-L}$, which is the subgroup of
$SU(4)_C$.

Furthermore, we split the $c$ stack of D6-branes into $c_1$ and
$c_2$ stacks with 2 D6-branes for each one. Similarly,  the gauge
fields and three multiplets in adjoint representation for
$SU(2)_R$ are broken down to respectively the gauge fields and
three multiplets in adjoint representation for $U(1)_{I_{3R}}$,
The $n_{\Ysymm}^c$ multiplets in symmetric representation for
$SU(2)_R$ are broken down to the $n_{\Ysymm}^c$ multiplets in
symmetric representation for $U(1)_{I_{3R}}$, while the
$n_{\Yasymm}^c$ multiplets in anti-symmetric representation for
$SU(2)_R$ are gone. In addition, there are $I_{c_1 c'_2}$ new
fields which are neutral under $U(1)_{I_{3R}}$ from the open
strings at the intersections of D6-brane stacks $c_1$ and $c_2'$.
And the rest of the particle spectrum is the same. Moreover, the
anomaly free gauge symmetry from $c_1$ and $c_2$ stacks of
D6-branes is $U(1)_{I_{3R}}$, which is the subgroup of $SU(2)_R$.

After D6-brane splittings, we obtain that the gauge symmetry is
$SU(3)_C\times SU(2)_L\times U(1)_{B-L} \times U(1)_{I_{3R}}$. To
break this gauge symmetry down to the SM gauge symmetry, we assume
that the
 minimal distance squared $Z^2_{(a_2 c_1')}$
 between the $a_2$ and $c_1'$ stacks of D6-branes on the third two-torus
is very small, then, we obtain $I_{a_2 c_1'}^{(2)}$ (the
intersection numbers for $a_2$ and $c_1'$ stacks on the first two
two-tori)
 pairs of chiral multiplets
with quantum numbers $({\bf { 1}, 1, -1, 1/2})$ and $({\bf { 1},
1, 1, -1/2})$ under $SU(3)_C\times SU(2)_L\times U(1)_{B-L} \times
U(1)_{I_{3R}}$
 from the light open string
states which stretch between the $a_2$ and $c_1'$ stacks of
D6-branes. These particles can break the $SU(3)_C\times
SU(2)_L\times U(1)_{B-L} \times U(1)_{I_{3R}}$ down to the SM
gauge symmetry and keep the D- and F-flatness
 because their quantum numbers are the same as those for the right-handed neutrino and its
complex conjugate. Especially, these particles are massless if
$Z^2_{(a_2c_1')}=0$.
 In summary, the complete symmetry breaking chains are
\begin{eqnarray}
SU(4)\times SU(2)_L \times SU(2)_R  &&
\overrightarrow{\;a\rightarrow a_1+a_2\;}\;  SU(3)_C\times SU(2)_L
\times SU(2)_R \times U(1)_{B-L} \nonumber\\&&
 \overrightarrow{\; c\rightarrow c_1+c_2 \;} \; SU(3)_C\times SU(2)_L\times
U(1)_{I_{3R}}\times U(1)_{B-L} \nonumber\\&&
 \overrightarrow{\;\rm Higgs \;
Mechanism\;} \; SU(3)_C\times SU(2)_L\times U(1)_Y~.~\,
\end{eqnarray}

The dynamical supersymmetry breaking in D6-brane models from Type
IIA orientifolds has been addressed in~\cite{CLW}. In the D6-brane
models, there are some filler branes carrying $USp$ groups which
are confining, and thus could allow  for gaugino condensation,
supersymmetry breaking and moduli stabilization.

The gauge kinetic function for a generic stack $x$ of D6-branes is
of the form (see, e.g., ~\cite{CLW}):
\begin{eqnarray}
f_x =  {\bf \textstyle{1\over 4}} \left[ n^1_x n^2_x n^3_x S -
(\sum_{i=1}^3 2^{-\beta_j-\beta_k}n^i_x l^j_x l^k_x U^i)  \right]
,\,
\end{eqnarray}
where the real parts of dilaton $S$ and moduli $U^i$ are
\begin{eqnarray}
{\rm Re}(S) = \frac{M_s^3 R_1^{1} R_1^{2} R_1^{3} }{2\pi g_{s}}~,~\, \\
{\rm Re}(U^{i}) = {\rm Re}(S)~ \chi_j \chi_k~,~\,
\end{eqnarray}
where $i\neq j\neq k$,  and $g_s$ is the string coupling. Also,
the K\"ahler potential is
\begin{eqnarray}
K=-{\rm ln}(S+ \bar S) - \sum_{I=1}^3 {\rm ln}(U^I +{\bar
U}^I).~\,
\end{eqnarray}
In our models, three stacks of D6-branes with $U(4)_C\times U(2)_L
\times U(2)_R$ gauge symmetry generically fix the complex
structure moduli $\chi_1$, $\chi_2$ and $\chi_3$ due to
supersymmetry conditions. So, there is only one independent
modulus field. To stabilize the modulus, we need at least two
$USp$ groups which are confining,
 {\it i.e.}, their $\beta$
functions are negative,
 and thus allow for
gaugino condensations~\cite{Taylor,RBPJS,BDCCM}. Suppose there are
$2N^{(i)}$ filler branes which are on top of $i$-th O6-plane and
carry $USp(N^{(i)})$ group. Its beta function is
\begin{eqnarray}
\beta_i^g&=&-3({N^{(i)}\over2}+1)+2 |I_{ai}|+
|I_{bi}| +  |I_{ci}|
+3({N^{(i)}\over2}-1)\nonumber\\
        &=&-6+2 |I_{ai}|+  |I_{bi}|+  |I_{ci}|~,~\,
\label{betafun}
\end{eqnarray}
where $3(N^{(i)}/2-1)$ is a contribution from three multiplets in
the anti-symmetric representation of the $USp$ branes. The
negative $\beta$ functions of $USp$ groups give strong constraints
on the intersection numbers of the associated filler branes and
the observable branes, and thus constrain the number of allowed
models.

If  supersymmetry turns out to  be broken due to the gaugino
condensations, the supersymmetry breaking will be mediated via
gauge interactions because the gravity mediated supersymmetry
breaking is much smaller. So, the supersymmetry CP problem can be
solved in our models. In this paper, we will neither study the
stabilization of this complex structure modulus (or for that
reason also  K\"ahler moduli, which could enter the gauge coupling
corrections due to the one-loop threshold corrections) due to
gaugino condensation,  nor the   issue of supersymmetry breaking
and postpone this for further study.
However, since we do eventually want to address these issues, we
confine our search only to models with at least two $USp$ gauge
group factors with negative $\beta$ functions.

\section{Systematic Search for Supersymmetric Pati-Salam Models}

The basic properties for the models that we want to construct are
given in Section III. Let us summarize them here. There are three
stacks of D6-branes, $a$, $b$, and $c$ with number of D6-branes 8,
4, and 4, which give us the gauge symmetry $U(4)_C$, $U(2)_L$ and
$U(2)_R$, respectively. We require that their intersection numbers
satisfy Eqs. (\ref{E3LF}$-$\ref{E3RF}). In addition, to stabilize
the modulus and possibly break the supersymmetry,
 we require that at
least two $USp$ groups in the hidden sector have negative $\beta$
functions.

Our searching strategy is the following: first, analytically
exclude most of the parameter space for the D6-brane wrapping
numbers which can not give the models with above properties, and
then scan the rest parameter space by employing a computer
program. If no two-torus is tilted, we can not have the particle
spectra with odd families of the SM fermions. So, there are three
possibilities: one tilted two-torus, two tilted two-tori, and
three tilted two-tori. The complete searching shows no-go for the
last two possibilities. As for the first one, all solutions are
tabulated in the Appendix. The detailed discussions are given in
the following three subsections. People only interested in
phenomenology may safely skip these parts.

\subsection{One Tilted Two-Torus}
Without loss of generality, let us suppose that the third
two-torus is tilted. Then, we may consider the two cases where the
$a$ stack of D6-branes is of NZ- and Z-type, which are
characterized by no and one zero wrapping number respectively. \\

Case I: NZ-type $a$ Stack of D6-branes with One Negative Wrapping Number \\

For NZ-type $a$ stack of D6-branes, among $A_a$, $B_a$, $C_a$ and
$D_a$, there is at least one equal to $-1$ in order to avoid the
tadpole cancellation condition (TCC) violations. Due to T-duality,
we may assume that $D_a=-1$. Obviously, the setup $I_{ab}=1$ or 2
can not be realized in this case. As for $|l_a^3|$ there are only
two possible absolute values (ABS): 1 and 3, because the third
wrapping number pairs (WNP) should be responsible for the even
factors of both $2^kI_{ac}=\pm 6$ and $2^kI_{ac'}=0$ ($k=1$). We
shall discuss this case according to the number of negative
wrapping numbers for a-brane because this number can not be larger
than 3 due to the D6-brane Sign Equivalent Principle (DSEP). Next,
let us take a look at the case with one negative wrapping number
first.

Since $D_a=-1$, the minus sign can only come from $l_a^1$, $l_a^2$
and $n_a^3$. Noticing that SUSY conditions can not be satisfied if
$n_a^3<0$, we set $l_a^1<0$ without loss of generality. If
$l_a^3=3$, the a-brane will be
\begin{eqnarray}
(+,-1)\times(+,1)\times(1,3)~.~\,
\end{eqnarray}
For the general intersection numbers $2^kI_{ax}=\pm 3\times 2^k$
and $I'_{ax}=0$, in order to generate the associated even factor
requires at least one set of WNPs from the tilted two-torus which
satisfy $|n_a^i|=|n_x^i|$, $|l_a^i|=|l_x^i|$, and
$n_a^il_a^i=-n_x^il_x^i=\pm3$ or $\pm1$ while the co-prime
conditions are also implemented. As a result, the 3rd WNP of the
C-type brane should be $(\pm1,\mp3)$, and thus one C2-brane is
needed for CTCC (Tadpole Cancellation Condition related to C)
requirement
\begin{eqnarray}
(-,+)\times(+,+)\times(-1,3)~.~\,
\end{eqnarray}
As for the third brane, if it is of A2-, B2- or D1-type,
\begin{eqnarray}
(-,+)\times(-,+)\times(-1,3)~,~ \\
(+,+)\times(-,+)\times(-1,3)~,~ \\
(+,+)\times(+,+)\times(1,-3)~,~\,
\end{eqnarray}
there will be a problematic intersection factor generated from the
second or the first WNP. This factor has a ABS larger than 1,
which will lead to $2^kI_{ax}>6$. If the third brane is of C2-type
also ATCC and DTCC require all wrapping numbers of a-, b- and
c-branes have unit ABS, which obviously is forbidden in our model
building. Therefore, the third brane is of Z-type and can not
provide extra positive Tadpole Charges. A direct result for this
is that $|n_a^1n_a^2|=1$ and $|n_C^1n_C^2|\leq 2$,
$|l_C^1l_C^2|\leq 2$ due to ATCC and DTCC. But, any one of
$|n_C^1n_C^2|$ and $|l_C^1l_C^2|$ can not be equal to 1 to avoid
vanishing $I_{aC}$, which may lead to $n_3^3=0$ due to ATCC and
DTCC again. This is impossible. Therefore, there is no solution
while $l_a^3=3$.

If $l_a^3=1$, the most general form for the a-brane is
\begin{eqnarray}
(L_1,-1)\times(L_2,1)\times(1,1)~,~\,
\end{eqnarray}
where $L_1$ and $L_2$ are positive. If $L_2 > 2$ for a-brane, no
matter what value $L_1$ has, both A2- and C2-type branes are
needed
\begin{eqnarray}
(-,+1)\times (-,+1)\times (-1,+1) ~,~\\
(-,+1)\times (+,+1)\times (-1,+1) ~.~\,
\end{eqnarray}
It is easy to see that the 2nd WNP of the A2-brane will contribute
an intersection factor larger than $L_2$ or 3. Since we have
another even factor from the third WNP, this may lead to
$2^kI_{aA}>6$. Therefore, there is no solution in this case.

For $L_2=2$, both A2- and C2-branes are still required. If both of
the 2nd and 3rd branes are of Z-type, to avoid ATCC and CTCC
violation, we must have $n_2^2=n_3^2=0$. Then the second WNPs of
these branes will yield an even intersection factor because $L_2$
is even, thus yielding even SM families. So, the 2nd and 3rd
branes can not be Z-type at the same time. If one of them is of
NZ-type, the other one must be NZ-type also because the
combination of a-brane and the second one will necessarily violate
ATCC and/or CTCC. As a result, b- and c-branes will be
\begin{eqnarray}
(-(L_1\pm1),+1)\times(-1,+1)\times(-1,+1)~,~\\
(-,+1)\times(+,+1)\times(-1,+1)~.~\,
\end{eqnarray}
It is obvious that the factor 3 for $2^kI_{aA}$ should be
generated from the second WNP, thus we have $|n_A^2l_A^2|=1$ here.
ATCC can not be satisfied while $L_1\geq 2$; as for $L_1=1$, ATCC
requires $|n_C^1n_C^2|\leq 2$, which will forbid that $I_{aC}$ be
a multiple of 3. Therefore, the only possible solution for a-brane
in this case is
\begin{eqnarray}
(L,-1)\times(1,1)\times(1,1) ~.~\,
\end{eqnarray}

Let us consider the case where $L>2$ first. In this case, one A2-
brane is required, and C- and D-type $USp$ groups can not appear
in the hidden sector since they are not asymptotically free
according to Eq. (\ref{betafun}). Noticing the intersection factor
3 of $2^kI_{aA}$ should be generated by the second WNP, we have
two kinds of possible solutions
\begin{eqnarray}
(L,-1)\times(1,1)\times(1,1) ~,~\\
(-(L\pm1),1)\times(-2,1)\times(-1,1) ~,~\,
\end{eqnarray}
and
\begin{eqnarray}
(L,-1)\times(1,1)\times(1,1) ~,~\,\\
(-(L\pm1),1)\times(-1,2)\times(-1,1) ~,~\,\\
(+,+)\times(-1,0)\times(-1,1) ~.~\,
\end{eqnarray}
For the former, no matter what the third brane is, the combination
of a- and A-branes yield no solution for moduli consistent with
supersymmetry; for the latter, obviously, a problematic
intersection factor larger than 3 comes out from the first WNPs of
a- and the third branes.

Finally, let us prove that the ABSs of b- and c-branes' wrapping
number can not be larger than 8 for $L \leq 2$. If they are
NZ-type branes, this conclusion is obvious since for each WNP only
one component's ABS can be larger than 2 to avoid TCC violation.
Thus the possible largest ABS of wrapping numbers should be less
than $2L+3=7$. For Z-type brane, the possible ABS larger than 8
should come from the WNP without zero wrapping number. Meanwhile,
the corresponding  two-torus is untilted. Let us focus on such a
WNP. Suppose this brane contributes the non-vanishing X- and
Y-type tadpole charges. If X- and Y- do not match with the types
of a- and the 3rd brane, due to the same reason applied to NZ-type
brane, the wrapping numbers' ABS of this brane still can not be
larger than 7. If $X=B$ and the 3rd brane is Y-type brane, we
shall have $l_Z^1n_Z^2=0$. For $l_Z^1=0$, in order to avoid the
problematic factor generated from the 2nd WNPs, we must have
$|n_Z^2l_Z^2|=2$ since $n_Z^2$ and $l_Z^2$ have different signs
due to SUSY conditions. For $n_Z^2=0$, given $|B_Z| \leq 2L+4\leq
8$ or $|n_Z^1| \leq 2L+4\leq 8$, $l_Z^1$ can not be larger than 7
due to the co-prime conditions for both $L=1$
and 2. \\

Case II: NZ-type $a$ Stack of D6-branes with Two
and Three Negative Wrapping Numbers \\

For the case with two minus signs, if they are from different
two-tori, the only possible setup for a-brane due to SUSY
conditions is
\begin{eqnarray}
(L_1,1)\times(L_2,-1)\times(1,-) ~,~\,
\end{eqnarray}
which corresponds to case I by the type II T-duality inference.
Therefore, we need not consider this case any more.

If the two minus signs come from the same two-torus, based on
DSEP, let us suppose that they are from the first two-torus. While
$l_a^3=3$, the a-brane will be of the form
\begin{eqnarray}
(-L_1,-1)\times(L_2,1)\times(1,3) ~,~\,
\end{eqnarray}
and both B2- and C2-branes
\begin{eqnarray}
(+,+)\times(-,+)\times(-1,3) ~,~\,\\
(-,+)\times(+,+)\times(-1,3) ~,~\,
\end{eqnarray}
are required. Obviously, a problematic intersection factor larger
than 1 will be generated for $2^kI_{aC}$. Therefore, there is no
solution at this time.

If $l_a^3=1$, the a-brane will be of the form
\begin{eqnarray}
(-L_1,-1)\times(L_2,1)\times(1,1) ~.~\,
\end{eqnarray}
For $L_1 \geq 2$ and $L_2 > 2$ or $L_1 > 2$ and $L_2 \geq 2$, B2-
and C2-branes are needed
\begin{eqnarray}
(+,+1)\times (-,+1)\times (-1,+1) ~,~\,\\
(-,+1)\times (+,+1)\times (-1,+1) ~,~\,
\end{eqnarray}
where at least one problematic intersection factor will be
contributed by the 2nd WNP of B2-brane or the 1st WNP of C2-brane.
Therefore the only possible solution for a-brane will be
\begin{eqnarray}
(-2,-1)\times(2,1)\times(1,1) ~,~\,
\end{eqnarray}
and
\begin{eqnarray}
(-L,-1)\times(1,1)\times(1,1) ~.~\,
\end{eqnarray}
Noticing that the second setup can be transformed to the setup in
Case I through type II T-duality, we only need to consider the
first one.

For the first setup, B2- and C2-branes are required and the only
possible solutions are
\begin{eqnarray}
(-2,-1)\times(2,1)\times(1,1) ~,~\,\\
(-1,1)\times(2\pm1,1)\times(-1,1) ~,~\,\\
(2\mp1,1)\times(-1,1)\times(-1,1) ~,~\,
\end{eqnarray}
Regretfully, these solutions are excluded because there are no
moduli solutions for supersymmetric D6-branes configuration. And
thus, the case with NZ-type $a$ stack of D6-branes with two
negative wrapping numbers
 should be ruled out.
For the case with three negative wrapping numbers,
 due to SUSY conditions the only
possible setup for a-brane is
\begin{eqnarray}
(L_1,-1)\times(L_2,-1)\times(-1,1)  ~,~\,
\end{eqnarray}
which obviously corresponds to the case with two minus signs
according to the type II T-duality variation. Thus, this case can
be excluded, too.\\

Case III: Z-type $a$ Stack of D6-branes \\

Now let us consider the Z-type a-brane. According to the type I
T-duality and its extended version, there are two possible setups
for a-brane: $A_a=B_a=0$ and $A_a=D_a=0$. The latter can not give
the required intersection numbers with b- and c-branes. If
$|n_a^3l_a^3|=3$, according to SUSY conditions, the a-brane has
four kinds of possible setups
\begin{eqnarray}
(0,-1)\times(L_1,L_2)\times(1,3) ~,~\,\\
(0,-1)\times(L_1,L_2)\times(3,1) ~,~\,\\
(0,1)\times(L_1,-L_2)\times(1,-3) ~,~\,\\
(0,1)\times(L_1,-L_2)\times(3,-1) ~.~\,
\end{eqnarray}
The 1st and 2nd setups correspond to the 4th and 3rd ones due to
type I T-duality, respectively, and the 1st one also corresponds
to the 3rd one due to type II T-duality. Therefore we only need to
consider the 1st one. For the 1st one, one C2-brane is required
and it has the form
\begin{eqnarray}
(-1,+)\times(+,1)\times(-1,3) ~.~\,
\end{eqnarray}
In order to avoid BTCC and DTCC violation (since the third brane
can provide at most one kind of positive tadpole charge), we must
have $l_C^2=1$. In addition, since $n_3^1\neq 0$ due to
$I_{a3}\neq 0$ , in order to avoid the BTCC violation we have to
require that the third brane satisfy $l_3^2=0$
\begin{eqnarray}
(-,-)\times(+,0)\times(-1,+) \label{1} ~.~\,
\end{eqnarray}
If $l_3^2\neq 0$, we must have $l_3^3=3$ and thus the third brane
is of B-type
\begin{eqnarray}
(+,+)\times(-,+)\times(-1,3) ~.~\,
\end{eqnarray}
A problematic intersection factor will arise from the second WNP.
Eq.~(\ref{1}) implies that the third brane is the b-brane and thus
c-brane is of C-type. Note that here $I_{ac}+I_{ac'}=-3$, which is
T-dual (type II) to the 3rd possible setup of a-brane with
$I_{ab}+I_{ab'}=3$, can not be achieved here. Since DTCC requires
$l_a^2=1$ and $|l_c^1|\leq 2$, $I_{ac}=-3$ will yield
$n_a^2=n_c^2+1>0$, and thus CTCC can not be satisfied. So, there
is no solution while $l_a^3=3$.

For the case with $|n_a^3l_a^3|=1$, based on the type I T-duality
and SUSY conditions, we can have only a-brane of the form:
\begin{eqnarray}
(0,-1)\times(L_1>0,L_2>0)\times(1,1) ~.~\,
\end{eqnarray}
Meanwhile, due to type II T-duality, we only need consider the
case $L_1>L_2$. Let us prove first that there is no solution while
$L_1>2$. If $L_1>2$, one C2-brane is required, but the third brane
must be of Z-type. The reason is that, if the third brane is of
A2- or B2-type
\begin{eqnarray}
(-,+)\times(-,+)\times(-1,1) ~,~\,\\
(+,+)\times(-,+)\times(-1,1) ~,~\,
\end{eqnarray}
there is a problematic intersection factor larger than $L_1$
generated from the second WNPs. If it is C2- or D1-brane,
\begin{eqnarray}
(-,+)\times(+,+)\times(-1,1) ~,~\, \\
(-,-)\times(+,+)\times(-1,1) ~,~\,
\end{eqnarray}
$n_3^2=l_3^2L_1+1$ or $l_3^2L_1+3$ is forbidden in order to avoid
the ATCC violation. Therefore, $I_{ab}$ and $I_{ac}$ have the same
signs if both of them are of C2- or D1-type, which is not allowed.
Thus $L_2=1$ is required.

Let us return to the C2-brane. If the intersection factor 3 for
$I_{aC}$ comes from the first two-torus, one additional A2-brane
is needed because $n_C^1=-3$ and $n_C^2\geq 2$ has led to ATCC
violation. Therefore, there is no solution in this case. If the
factor 3 comes from the second two-torus, $l_C^2$ can not be
larger than 2, otherwise, DTCC can not be satisfied. The only two
possible solution for a- and C-type branes will be
\begin{eqnarray}
(0,-1)\times(3,1)\times(1,1) ~,~\, \\
(-1,+)\times(3,2)\times(-1,1) ~,~\,
\end{eqnarray}
and
\begin{eqnarray}
(0,-1)\times(L,1)\times(1,1) ~,~\, \\
(-1,+)\times(L-3,1)\times(-1,1) ~.~\,
\end{eqnarray}
For the first possible solution, because DTC (ATC, BTC, DTC and
DTC denote the A-, B-, C- and D-type tadpole charges,
respectively.) is already filled, the 3rd brane should satisfy
$l_3^1l_3^2=0$ and have the form
\begin{eqnarray}
(-1,l_3^1)\times(1,l_3^2)\times(-1,1) ~,~\,
\end{eqnarray}
where $A_3=-1$ is required by ATCC. Obviously, the intersection
factor 3 can not be generated for $I_{a3}$. For the second
possible solution, ATCC requires $4\leq L\leq7$. To avoid the CTCC
violation, $l_C^1>2$ is necessary, which, however, will lead to
the DTCC violation. As a result, there is no solution if $L_1>2$.

For $L=1, 2$, a solution is possible. Here we shall prove that the
wrapping numbers of b- and c-branes can not be larger than 8. For
a NZ-type brane, the only possible wrapping number with its ABS
larger than 3 is $l^1$. $|n^2|$ and $|l^2|$ must be smaller than
$2L+3$ since the smaller one still can not be larger than 2.
$|l^1|$ can be larger than 3 only when the NZ-brane is of C- or
D-type. Correspondingly, the third brane should be of D- or
C-type. Since the components of the 3rd WNPs for both branes have
different signs, the two branes should be of C2- and D1-types
\begin{eqnarray}
(0,-1)\times(L,1)\times(1,1) ~,~\, \\
(+,l_C^1<0)\times(+,+)\times(1,-1) ~,~\, \\
(+,+)\times(+,+)\times(1,-1) ~.~\,
\end{eqnarray}
Due to ATCC and BTCC, one of $I_{aC}$ and $I_{aD}$ can not be a
multiple of 3. For a Z-type brane, as we discussed before, the
possible wrapping number with its ABS larger than 3 can only come
from an untilted two-torus and the corresponding WNP contains no
zero wrapping number. If this WNP is from the second torus, SUSY
conditions require that its two components have different signs
and thus their ABS can not be larger than 2 to avoid a problematic
intersection factor. If this WNP is from the first two-torus, the
only possible large wrapping number is $l_Z^1$. If $l_Z^1>2$, the
third brane is of C-type or D-type. For the C-type case, given
that ATCC and DTCC require $|A_3|\leq 3$ and $|D_3|\leq 2$,
respectively, CTCC will yield $|C_Z|$ or $|l_Z^1|\leq
|A_3|\times|D_3|+4-2=8$. As a result, the largest ABS of these
wrapping numbers is less than 9. We shall have the same situation
for
D-type 3rd brane.\\

\subsection{Two Tilted Two-Tori}
Similar to the previous Subsection, we consider the NZ-type $a$
stack of D6-branes first and then turn to the Z-type one.
Here we assume that the second and the third two-tori are tilted.\\

Case I: NZ-type $a$ Stack of D6-branes with One Negative Wrapping Number \\

For the a-brane of NZ-type, let us still suppose $D=-1$ in
accordance with T-duality. Then this brane has two forms
\begin{eqnarray}
(L_1,-1)\times(L_2,1)\times(1,L_3) ~,~\,
\end{eqnarray}
and
\begin{eqnarray}
(L_1,1)\times(L_2,-1)\times(1,L_3) ~,~\,
\end{eqnarray}
where $L_1$, $L_2$, $L_3$ are positive, and $L_2$, $L_3$ are odd.
According to the requirement of moduli stabilization for gauge
symmetries in the hidden sector, there is at most one with its ABS
larger than 2 among all wrapping numbers of a-brane, which implies
three possibilities: (1)$L_3=1$, $L_1\leq 2$ and $L_2\geq 3$;
(2)$L_2=1$, $L_1\leq 2$ and $L_3\geq 3$; (3)$L_2=L_3=1$ and
$L_1\geq 2$.

Now let us consider the first form of a-brane and set $L_2\geq 3$
\begin{eqnarray}
(L_1,-1)\times(L_2,1)\times(1,1) ~,~\,
\end{eqnarray}
then the other two branes are of A- and C-types, respectively, and
have $D_A=D_C=-1$ to avoid DTCC violation. Meanwhile, to avoid a
problematic intersection factor, the C-brane should be of C2-type:
\begin{eqnarray}
(-L_1\pm1<0,+1)\times(L_2\pm6>0,+1)\times(-1,1) ~,~\,
\end{eqnarray}
where the intersection factor 3 of $I_{aC}$ can not come from the
first WNP. If it does, we shall have $n_C^1=-(L_1+3)$ because of
$0<L_1\leq 2$. The C-type $USp$ group is no longer asymptotically
free, and then we do not have enough $USp$ groups in the hidden
sector to stabilize the modulus. Now let us consider the A-type
brane. For A1-brane, the 0 intersection factor for $I_{aA}$ can
not be generated. The A-brane, therefore, should be of A2-type.
The factor 3 of $I_{aA}$ comes from the second WNP, and thus the
brane will be
\begin{eqnarray}
(-L_1\pm1<0,+1)\times(L_2\pm6<0,+1)\times(-1,1) ~.~\,
\end{eqnarray}
Under this setup, we always have $n_A^1=n_C^1$ no matter $L_1$ is
equal to 1 or 2. This is obvious for $L_1=1$. If $L_1=2$, both
$n_A^1$ and $n_C^1$ can not be equal to 3. Otherwise, C-type $USp$
group is not asymptotically free again. So we shall have
$n_A^1=n_C^1=-1$. On the other hand, because of $n_C^2>0$ and
$n_A^2<0$, we must have $n_C^2=L_2+6$ and $n_A^2=L_2-6$, and thus
$|n_C^2|-|n_A^2|=2L_2>0$. Due to $L_2\geq 3$, the ATCC can not be
satisfied for any value of $L_2$. As a result, we have $L_2=1$ for
the first setup of a-brane.

If $L_3\geq 3$, the a-brane has the form
\begin{eqnarray}
(L_1,-1)\times(1,1)\times(1,L_3)~,~\,
\end{eqnarray}
and then one C-type brane is required due to CTCC. At this time A-
and D-type $USp$ groups are not asymptotically free due to the
large value of $L_3$. If the C-brane is of C1-type, the
intersection factor 3 for $I_{aC}$ should come from the 1st WNP
and the only possible solutions for the a- and C1-branes are
\begin{eqnarray}
(L_1,-1)\times(1,1)\times(1,L_3) ~,~\,\\
(n_C^1>0,l_C^1>0)\times(1,-1)\times(1,L_3+2) ~,~\,
\end{eqnarray}
where $L_1 n_C^1 l_C^1=2$ to provide the factor 3 for $I_{aC}$,
$n_C^3$ can not be equal to 2 to keep $ATC=DTC=0$, and $l_C^3$ can
not be equal to $L_3-2$ to avoid CTCC violation. Now it is easy to
check that there is no moduli solution while $L_1=2$ or $l_C^1=2$,
and B-type $USp$ group can not be generated in the hidden sector
while $n_C^1=2$. If this is a C2-brane, the intersection factor 3
will come from the third WNP, which means $L_3=3$, thus we shall
have $n_C^2l_C^2\geq 3$. Noticing that the third brane is of
NZ-type no matter which one is larger, it is easy to check that
there are no enough asymptotically free $USp$ groups available in
the hidden sector.

Finally, let us consider the last kind of setup of a-brane
\begin{eqnarray}
(L_1\geq 2,-1)\times(1,1)\times(1,1) ~.~\,
\end{eqnarray}
For $L_1\geq 3$, one brane of A-type is required. This is also
true for $L_1=2$. Otherwise, to avoid one A-type brane, the fact
that ATC is full will lead to $n^1n^2n^3=0$ for the other two
branes. $n^1$ can not be equal to zero to avoid an even
problematic intersection factor. If $n^2=0$, the intersection
factor 3 comes from the first WNP and this brane has the form
\begin{eqnarray}
(-2l^1\pm3<0,l^1)\times(0,2)\times(-1,1) ~.~\,
\end{eqnarray}
To avoid the DTCC violation, the third brane must have $n_3^3=0$
and $|l_3^3|=2$. This indicates that no enough $USp$ groups are
available in the hidden sector now. As a result, one A-type brane
is definitely necessary while $L_1\geq 2$.

This A-type brane is a A2-brane to get $I_{aA'}=0$. If the
intersection factor 3 of $I_{aA}$ comes from the first WNP, the
possible solution for this brane is
\begin{eqnarray}
(-(l_A^1L_1\pm 3)<0,l_A^1)\times(-1,1)\times(-1,1) ~,~\,
\end{eqnarray}
which gives out no moduli solutions by combining with a-brane
setup. If the factor 3 is given out by the second WNPs, we must
have $L_1=2$, in order to preserve at least two asymptotically
free $USp$ groups in the hidden sector. Thus the possible solution
for a- and A-type brane will be
\begin{eqnarray}
(2,-1)\times(1,1)\times(1,1) ~,~\, \\
(-1,1)\times(-(6-l_A^2)<0,l_A^2)\times(-1,1) ~,~\,
\end{eqnarray}
which yield no moduli solutions while $l_A^2<3$. If $l_A^2=4$ or
5, DTCC requires that the third brane be of D-type, which
necessarily leads to less than two asymptotically free $USp$
groups available in the hidden sector.

As for the second form of a-brane in this case,
\begin{eqnarray}
(L_1,1)\times(L_2,-1)\times(1,L_3) ~,~\,
\end{eqnarray}
where $L_1\geq 2$, $L_2\geq 3$ and $L_3\geq 3$, the derivation is
similar. For the case $L_1\geq 2$, both A- and B-type branes are
required. This is obvious for $L_1\geq 3$. For $L_1=2$, if there
is no A- and B-type branes, the fact that ATC and BTC are filled
requires $n_2^1=n_3^1=0$, in order to avoid the ATCC and BTCC
violation, which in turn will lead to a problematic even factor
from the first WNPs since $L_1$ is even. These two necessary
NZ-type branes have $D_A=D_B=-1$ due to DTCC. If the B-brane is of
B1-type, the factor 3 for $I_{aB}$ comes from the first WNP and
thus we have $n_a^1=2$ and $L_2=1$. Then, the possible solutions
for a-brane and the B-brane will be
\begin{eqnarray}
(2,1)\times(1,-1)\times(1,1) ~,~\,\\
(1,-1)\times(1,1)\times(1,3) ~,~\,
\end{eqnarray}
with $L_3=1$ required by the BTCC because of $B_B \leq L_3+2$. It
is easy to check that under this setup, A1-brane can not obtain
the intersection factor 3 to avoid the BTCC and CTCC violation,
and A2-brane can not yield the moduli solution by combining with
the B-type brane.

Now let us consider the combination of B2- and A1-branes. $L_2$
can not be larger than 3, otherwise, one problematic factor will
be generated for $I_{aA}$ from the second WNPs. If $L_2=3$, B- and
D-type $USp$ groups are no longer asymptotically free. To ensure
C-type $USp$ group is asymptotically free, we must have $L_1=2$.
Then the only possible solutions are
\begin{eqnarray}
(2,1)\times(3,-1)\times(1,1) ~,~\, \\
(+,1)\times(-,1)\times(-1,1) ~,~\, \\
(-1,-1)\times(3,1)\times(1,3) ~.~\,
\end{eqnarray}
But, due to BTCC, $n_B^1$ can not be smaller than 3, $i.e.$,
C-type $USp$ group is still not asymptotically free. So we must
have $L_2=1$. As for $L_3$, we shall have the same situation if it
is larger than 1. So, the last possible solutions for a-, B- and
A-branes are
\begin{eqnarray}
(L_1,1)\times(1,-1)\times(1,1) ~,~\, \\
(L_1\pm3>0,1)\times(-3,1)\times(-1,1) ~,~\, \\
(-(L_1\pm3)<0,-1)\times(1,1)\times(1,3) ~,~\,
\end{eqnarray}
where $l_A^3=-n_B^2=3$ is because the intersection factor 3s of
$I_{aB}$ and $I_{aA}$ cannot come from the tilted two-tori. This
is straightforward to see. For example, if the intersection factor
3 of $I_{aB}$ comes from the 2nd WNP, we must have $n_B^2=-7$ and
then the ATCC will require $n_A^1\geq 7$. This is not allowed
since B-, C- and D-type $USp$ groups are not asymptotically free
in this case. For this set of possible solutions, ATCC requires
$L_1\leq 3$, then we shall have $n_B^1=-n_A^1=L_1+3$; BTCC is
violated. As for the B2-A2 combination, the factor 3 for $I_{aA}$
comes from the first WNP, and thus we have $L_3=1$. The possible
solutions are
\begin{eqnarray}
(2,1)\times(1,-1)\times(1,1) ~,~\, \\
(n_B^1>0,1)\times(3,-1)\times(1,-1) ~,~\, \\
(-1,1)\times(-3,1)\times(-1,1) ~,~\,
\end{eqnarray}
where $L_2=1$ is required by ATCC because of $A_A\leq L_2+2$, and
$n_B^2\leq 3$ is required by CTCC. Now it is easy to see that ATCC
will forbid $I_{aB}$ obtain the intersection factor 3. As a
result, for $L_1\geq 2$, there is no solution.

We now turn to $L_1=1$ case. If $L_2\geq3$, B- and D-type $USp$
groups can not appear in the hidden sector and a-brane will have
the form
\begin{eqnarray}
(1,1)\times(L_2,-1)\times(1,1)  ~.~\,
\end{eqnarray}
One A-type brane is required. For A1-brane, the factor 3 for
$I_{aA}$ will be contributed by the second WNPs and thus we have
$n_A^3l_A^3=3$. If $n_A^3=3$, the third brane is of D1-type since
D2-type brane can not generate the zero factor for $I_{aD'}$.
Because no extra positive BTC is available, we must have
$B_a=B_A=B_D=1$. Then the only possible solutions are
\begin{eqnarray}
(1,1)\times(3,-1)\times(1,1) ~,~\, \\
(-1,-2)\times(3,1)\times(3,1) ~,~\, \\
(1,2)\times(3,1)\times(1,-1) ~,~\,
\end{eqnarray}
which, however, is excluded by CTCC. If $l_A^3=3$, the third brane
is of B-type and it should be B2-brane since the problematic
intersection factor can not be avoided for B1-brane. Due to
$BTC=DTC=0$, the possible solutions are
\begin{eqnarray}
(1,1)\times(3,-1)\times(1,1) ~,~\, \\
(-2,-1)\times(3,1)\times(1,3) ~,~\, \\
(4,1)\times(1,-1)\times(1,-1) ~,~\,
\end{eqnarray}
which means that the C-type $USp$ group is not asymptotically
free, either.

If the A-brane is of A2-type, the factor 3 of $I_{aA}$ will be
provided by the first WNPs, indicating $|n_A^1l_A^1|=2$.
Meanwhile, $l_A^2$ can not be larger than 1 to satisfy $BTC=DTC=0$
while the 3rd brane is also counted in. Then the possible solution
for a- and A2-brane will be
\begin{eqnarray}
(1,1)\times(3,-1)\times(1,1) ~,~\, \\
(-1,2)\times(n_A^2<0,1)\times(-1,1) ~,~\,
\end{eqnarray}
where $l_A^1=2$ is because there is no moduli solutions for
$l_A^1=1$. But, no proper value is available for $n_A^2$:
$|n_A^2|=1$ will lead to ATCC or DTCC violation when the third
brane is included, and $|n_A^2|=5$ will make C-type $USp$ group
unavailable since the third brane must be of Z-type to satisfy
$DTC=0$. As a result, there is no solution for $L_2\geq 3$. As for
the case where $L_3\geq 3$, due to the extended type I T-duality,
it is dual to the case where $L_2\geq 3$ according to Eq.
(\ref{T-duality Ia}), and thus all above conclusion can be applied
directly to this case.

At last, let us recall that the ABS of all branes' wrapping
numbers are less than 8 under the survived two setups of a-brane
\begin{eqnarray}
(1,-1)\times(1,1)\times(1,1) ~,~\, \\
(1,1)\times(1,-1)\times(1,1) ~.~\,
\end{eqnarray}
For the former, the proof is exactly the same as in the case of
one-tilted two-torus. For the latter, the only difference happens
where $n_2^1=0$ and the third brane is of D-type. In that case,
ATCC, BTCC and CTCC will impose on the 3rd brane the constraint
$D_3\leq5$. Combining this constraint and CTCC, it implies that
the wrapping numbers of the 2nd brane can not be
larger than 8 in this case, either.\\

Case II: NZ-type $a$ Stack of D6-branes with Two
and Three Negative Wrapping Numbers \\

In this case, the forms of a-brane allowed by the SUSY conditions
include
\begin{eqnarray}
(-L_1,-1)\times(L_2,1)\times(1,L_3)~,& (L_1,-1)\times(L_2,1)\times(1,-L_3)~, \\
(L_1,1)\times(L_2,-1)\times(1,-L_3)~,&
(-L_1,1)\times(L_2,-1)\times(1,-L_3) ~.~\,
\end{eqnarray}
The first case corresponds to
\begin{eqnarray}
(L_1,-1)\times(L_3,1)\times(1,L_2)  ~,~\,
\end{eqnarray}
by a T-duality combination of the Eqs. (\ref{T-duality Ia}) and
(\ref{T-duality IIa}). As for the last three cases, they
correspond to
\begin{eqnarray}
(L_1,1)\times(L_2,-1)\times(1,L_3) ~,~\, \\
(L_1,-1)\times(L_2,1)\times(1,L_3) ~,~\, \\
(-L_1,-1)\times(L_2,1)\times(1,L_3) ~,~\,
\end{eqnarray}
respectively, according to the extended type II T-duality Eq.
(\ref{T-duality IIa}). Therefore, these setups are also excluded
due to T-duality.\\

Case III: Z-type $a$ Stack of D6-branes \\

Let us consider the case where the a-brane is of Z-type. Suppose
that the zero wrapping number comes from the untilted two-torus,
then due to T-duality, we can assume $A_a=B_a=0$ or $n_a^1=0$ and
this a-brane has the form
\begin{eqnarray}
(0,-1)\times(+,+)\times(+,+)  ~.~\,
\end{eqnarray}
For the four unknown positive wrapping numbers, at least one of
them is larger than 2; if not, all of them are equal to 1. Without
loss of generality, we may suppose $n_a^2$ is such a wrapping
number, then it must be true that $l_a^2=n_a^3=l_a^3=1$. Before
verifying this, we point out first that the other two branes
should be of C- and D-type only if one of $l_a^2$, $n_a^3$ and
$l_a^3$ is not equal to 1. This is obvious if $l_a^3$ is not the
largest one among these three wrapping numbers. If $l_a^3$ is the
largest one, it should be larger than 2, and then A- and B-type
$USp$ groups are no longer asymptotically free. C-type brane is
definitely required due to $n_a^2\geq 3$ here. So, to preserve the
D-type $USp$ group in the hidden sector, we must have one D-type
brane to provide the extra positive DTC. Next, we shall discuss
these cases separately.

For $n_a^3>1$, the intersection factor 3s of $I_{aC}$ and $I_{aD}$
will come from the tilted two-tori, which means, among $n_C^2$,
$n_C^3$, $n_D^2$ and $n_D^3$, there are two whose ABS is equal to
3. This will lead to ATCC violation. Thus we must have $n_a^3=1$.
For $l_a^3$, it can not be larger than 3, otherwise, we shall have
$n_C^2=n_D^2=3$ to generate the intersection factor 3 and ATCC
will be violated again. If $l_a^3=3$, the only possible solutions
are
\begin{eqnarray}
(0,-1)\times(3,1)\times(1,3) ~,~\, \\
(1,l_2^1)\times(3,-1)\times(1,1) ~,~\, \\
(1,l_3^1)\times(1,1)\times(1,-3) ~.~\,
\end{eqnarray}
Obviously, no matter what values $l_2^1$ and $l_3^1$ take, CTCC
and DTCC can not be satisfied at the same time. So $l_a^3$ must be
equal to 1 also. For $l_a^2>1$, suppose that $n_a^2>l_a^2$ due to
T-duality, the intersection factor 3, no matter for  $I_{aC}$ or
$I_{aD}$, can not come from the first WNP. For example, if
$I_{aC}$'s does, ATCC and BTCC will require $A_D=B_D=-1$ since now
we have $A_C=B_C=-3$. Then the D-type brane can not obtain the
intersection factor 3. As a result, the only possible solutions
will be
\begin{eqnarray}
(0,-1)\times(n_a^2,l_a^2)\times(1,1) ~,~\, \\
(-1,l_C^1>0)\times(n_C^2>0,l_C^2>0)\times(-1,1) ~,~\, \\
(1,l_D^1>0)\times(n_D^2>0,l_D^2>0)\times(1,-1) ~.~\,
\end{eqnarray}
Since $|n_C^2l_C^2|\leq 3$ and $|n_D^2l_D^2|\leq 3$, it is not
hard to check one by one that the intersection factor 3 of
$I_{aC}$ and $I_{aD}$ can not be generated from the second WNPs at
the same time. So, the a-brane can not have two wrapping numbers
with their ABSes larger than 1 at the same time. The last
possibility is then
\begin{eqnarray}
(0,-1)\times(n_a^2>0,1)\times(1,1)  ~.~\,
\end{eqnarray}

For the last possible setup of a-brane, if $n_a^2\geq 3$, one
C-type brane is required. For a C1-brane, the intersection factor
3 for $I_{aC}$ comes from the second WNP, and the possible
solutions of a- and C1-branes are
\begin{eqnarray}
(0,-1)\times(3,1)\times(1,1) ~,~\, \\
(1,l_C^1>0)\times(3,-1)\times(1,3)  ~,~\,
\end{eqnarray}
where $n_C^3$ must be equal to 1 in order to avoid ATCC and/or
DTCC violation after the third brane is included. Meanwhile, to
generate the intersection factor 3 of $I_{a3}$, $|n_3^1n_3^2|\geq
3$ or $|A_3|\geq 3$ can not be avoided, so the last brane should
be of A-type. Noticing that $|n_3^1|=3$ is not allowed by BTCC, we
have $|n_3^2|=3$, which necessarily leads to no enough
asymptotically free $USp$ groups in the hidden sector. For a
C2-brane, the intersection factor 3 of $I_{aC}$ can not come from
the first WNPs. If it does, one extra D- or A-type brane is
required. For a D-type brane, $A_D=B_D=1$ will forbid $I_{aD}$ to
obtain the intersection factor 3. For an A1-brane, BTCC and DTCC
require $|n_A^3l_A^3|=1$ which will lead to $I_{aA}=0$. For an
A2-brane, we have $B_A=D_A=D_C=-1$ and thus the intersection
factor 3 of $I_{aA}$ can only come from the second WNPs. The only
possible solutions are
\begin{eqnarray}
(0,-1)\times(3,1)\times(1,1) ~,~\, \\
(-3,1)\times(n_C^2,1)\times(-1,1) ~,~\, \\
(-1,1)\times(-3,1)\times(-1,1)  ~.~\,
\end{eqnarray}
Obviously $n_C^2$ has no solution to satisfy CTCC and ATCC at the
same time. As a result, the intersection factor 3 of $I_{aC}$ has
to come from the 2nd WNPs and the only possible solutions for a-
and C2-brane will be
\begin{eqnarray}
(0,-1)\times(n_a^2\geq3,1)\times(1,1)  ~,~\, \\
(-1,l_C^1>0)\times(n_a^2\pm6>0,1)\times(-1,1) ~,~\,
\end{eqnarray}
where $l_C^2$ can not be larger than 3 to avoid the requirement of
additional B- and D-type branes. If it is equal to 3, there is no
solution for $n_a^2$ satisfying the co-prime condition. It can not
be equal to 2 or 3 also to avoid the requirement of additional A-
and D-type branes are required at the same time. If it is equal to
2, we have $n_C^2\geq 4$ and two additional ATC and DTC are
required together. As for the $n_C^2$, if it is equal to
$n_a^2+6$, the third brane is of A-type, and we have $B_A=-3$ and
$D_A=-1$, or $l_A^2=1$ and $|n_A^1l_A^3|=3$. If $l_A^3=3$, this is
a A1-brane and $n_A^2=n_a^2\pm 6>0$. It is easy to check that in
either case, ATCC and CTCC can not be satisfied at the same time.
If $l_A^3=1$, this is a A2-brane and we have $n_A^1=3$. At this
time, a problematic intersection factor for $I_{aA}$ will be
generated from the 2nd WNPs. If $n_C^2=n_a^2-6$, CTCC requires
$l_C^1>2$ and thus the 3rd brane is of D-type. Meanwhile, we have
$B_D=n_D^1l_D^2l_D^3=-3$. For a D2-type brane, one problematic
intersection factor for $I_{aD}$ will be generated by the 2nd WNPs
due to $n_a^2\geq 7$. For a D1-type brane, $l_D^3=-1$; if
$l_D^2=3$, $n_D^1=1$ and there is no solution for $n_D^2$
satisfying the co-prime condition. Therefore, the possible
solutions are
\begin{eqnarray}
(0,-1)\times(n_a^2\geq7,1)\times(1,1) ~,~\, \\
(-1,l_C^1>0)\times(1,1)\times(-1,1) ~,~\, \\
(3,l_D^1>0)\times(1,1)\times(1,-1) ~,~\,
\end{eqnarray}
regretfully, there is no solution for $l_C^1$ and $l_D^1$
satisfying CTCC and DTCC at the same time. Based on the above
analysis,
 we can declare the no-go theorem for $n_a^2>1$. As for the case
where $n_a^2=1$, following the same logic as applied in the case
of one-tilted two-torus, it is easy to see that no wrapping number
can have ABS larger than 8.

If the zero wrapping number comes from one of the tilted two-tori,
due to type I T-duality and its extended version again, we may
take $n_a^2=0$ and thus $|l_a^2|=2$. The other two branes will be
B2- and D1-branes. If they are of Z-type, their second WNPs have
to satisfy $|n^2|=2$ and $l^2=0$, which will yield a problematic
intersection factor 4 since the third WNPs have yielded another
even factor. Furthermore, they must be of B2- and D1-types because
B1- and D2-types branes can not provide the zero factor to
$I_{aB'}$ and $I_{aD'}$. The possible solutions then are
\begin{eqnarray}
(n_a^1>0,l_a^1>0)\times(0,-2)\times(1,1)  ~,~\, \\
(+,+)\times(1,-)\times(1,-1) ~,~\, \\
(+,+)\times(1,+)\times(1,-1) ~,~\,
\end{eqnarray}
up to the inferential Type II T-duality. Here the intersection
factor 3 for $I_{aB}$ and $I_{aD}$ can not come from the 2nd WNPs.
If one of them does obtain the intersection factor 3 from the 2nd
WNPs, the other one will be forbidden to obtain the intersection
factor 3 due to ATCC and CTCC constraints. The factor 3 can not
come from the 3rd WNPs also since one of ATCC and CTCC will be
violated only if $n_a^3$ or $l_a^3$ is equal to 3. Now let us
consider the last case -- they are from the 1st WNPs. To generate
these intersection factor 3, $n_a^1=l_a^1=1$ is not allowed, since
$n_B^1l_B^1\leq 3$ and $n_D^1l_D^1\leq 3$ due to ATCC and CTCC.
Without loss of generality, let us suppose $n_a^1<l_a^1$. Then
according to $I_{aB}=-I_{aD}$, we have
\begin{eqnarray}
(l_B^1+l_D^1)n_a^1=(n_B^1+n_D^1)l_a^1  ~.~\,
\end{eqnarray}
Given $l_B^1+l_D^1\leq 4$, $n_B^1+n_D^1\leq 4$ and the co-prime
conditions, $(n_a^1,l_a^1)$ have three possible solutions $(3,4)$,
$(2,3)$ and $(1,2)$. Now it is not hard to figure out one by one
that the intersection factor 3s can not be generated from the 1st
WNPs for $I_{aB}$ and $I_{aD}$ at the same time. As a result,
there is no solution in this case.

\subsection{Three Tilted Two-Tori}

If all two-tori are tilted, for one set of definite values
arranged to $\{A, B, C, D\}$, models characterized by any
permutations of them actually are T-dual to each other. If a-brane
is of NZ-type, considering that there is only one wrapping number
$L$ with its ABS larger than 1 to preserve enough asymptotically
free $USp$ groups, we only need to consider two kinds of possible
setups for a-brane: $\{A=-B=L, C=D=-1\}$ and $\{A=B=-L, C=-D=1\}$.
And the C- and D-type $USp$ groups are not asymptotically free.

For the former, without loss of generality, we may suppose that
a-brane is of A1-type up to the inferential type II T-duality,
\begin{eqnarray}
(-L,-1)\times(1,1)\times(1,1)  ~.~\,
\end{eqnarray}
Then one of the other two branes must be of B-type. In the case of
a B1-brane, the intersection factor 3 will come from the 1st WNP
and $L=n_B^1=n_B^2l_B^2=n_B^3l_B^3=3$. So, no enough
asymptotically free $USp$ groups can be preserved. In the case of
a B2-brane, still for the sake of enough asymptotically free $USp$
groups in the hidden sector, the intersection factor 3 must come
from the first WNPs. Meanwhile, due to CTCC and DTCC, we have
$l_B^1=1$ or $l_B^1=l_B^2=l_B^3=-n_B^2=-n_B^3=1$ where there are
no moduli solutions for a- and B2- branes.

For the latter, we may suppose that a-brane is a C1-brane
\begin{eqnarray}
(L,1)\times(1,-1)\times(1,1) ~.~\,
\end{eqnarray}
Then the other two branes must be of A- and B-types and
$D_A=D_B=-1$ since a-brane is of C-type. In order to keep enough
asymptotically free $USp$ groups, the intersection factor 3 must
come from the first WNPs and only the combinations of A1- and
B2-branes, and A2- and B1-branes are allowed
\begin{eqnarray}
(L,1)\times(1,-1)\times(1,1) ~,~\, \\
(-(L\pm6)<0,-1)\times(1,1)\times(1,3) ~,~\, \\
(L\pm6>0,1)\times(-3,1)\times(-1,1) ~,~\,
\end{eqnarray}
and
\begin{eqnarray}
(L,1)\times(1,-1)\times(1,1) ~,~\, \\
(-(L\pm6)<0,1)\times(-3,1)\times(-1,1) ~,~\, \\
(L\pm6>0,-1)\times(1,1)\times(1,3) ~.~\,
\end{eqnarray}
For the first one, ATCC and BTCC can not be satisfied only if
$L>1$; for the second one, the combination of A2- and B1-branes
gives no moduli solutions.

As for the case where $L=1$, it is easy to see that the ABS of all
branes' wrapping numbers are less than 8. Meanwhile, if a-brane
has one vanishing wrapping number, supposing $(0,-2)$ for the
second WNP of a-brane, we know that the intersection factor 3 of
$I_{aB}$ and $I_{aD}$ cannot come from the 2nd WNPs. Meanwhile,
they can not come from the same WNPs also due to ATCC and CTCC.
So, the only possible solutions are
\begin{eqnarray}
(1,3)\times(0,-2)\times(3,1) ~,~\, \\
(1,-3)\times(1,\pm|l_2^2|)\times(1,1) ~,~\, \\
(1,1)\times(1,\pm|l_3^2|)\times(3,-1) ~.~\,
\end{eqnarray}
However, in this case the BTCC or DTCC is violated. Thus, no
solution exists!

\subsection{Preliminary Phenomenological Features of the Models}

After analytically excluding  most of parameter space for wrapping
numbers, we wrote  a computer  program to scan the rest parameter
space. The results indicate that no model is available for the
cases with two and three tilted two-tori. For the case with
one-tilted two-torus, we obtain 11 inequivalent models which can
be specified by  two classes.

The first class include Tables \ref{model I-NZ-1a}$-$\ref{model
I-Z-8}, which correspond to the models that do not possess the
gauge coupling unification for $SU(2)_L$ and $SU(2)_R$  at the
string scale and whose Higgs doublets are generated by the
intersections of b- and c-stacks of branes.

For the second class, which includes Tables \ref{model I-Z-9} and
\ref{model I-Z-10}, they have the $SU(2)_L$ and $SU(2)_R$ gauge
coupling unification at the string scale. Because the b-stack
branes for these models are parallel to both c-stack branes and
their images on one of the three two-tori, their Higgs doublet
pairs come from the massless open string states in a $N=2$
subsector and   form vector-like pairs.

For all of these  models, except I-Z-7, I-Z-8, and I-Z-9, the
number of the pairs of Higgs doublets is less than 9, which could
make them phenomenologically interesting. In particular, there are
two pairs of Higgs doublets in model I-NZ-1a. As an example, the
chiral particle spectra for models I-NZ-1a, I-Z-5 and I-Z-10 are
presented in Tables \ref{spectrum I-NZ-1a}-\ref{spectrum I-Z-10},
respectively. For model I-NZ-1a, it has 2 pairs of Higgs doublets
and two  confining hidden sector gauge group factors. Model I-Z-5
has 3 pairs of Higgs doublets and three confining gauge group
factors.  Model I-Z-10 has the gauge coupling unification of
$SU(2)_L$ and $SU(2)_R$ at the string scale and 4  confining gauge
group factors.  These models thus possess potentially
phenomenologically attractive freatures.

\begin{table}
[htb] \footnotesize
\renewcommand{\arraystretch}{1.0}
\caption{The chiral spectrum in the open string sector of model
I-NZ-1a} \label{spectrum I-NZ-1a}
\begin{center}
\begin{tabular}{|c||c||c|c|c||c|c|c|}\hline
I-NZ-1a & $SU(4)\times SU(2)_L\times SU(2)_R $
& $Q_4$ & $Q_{2L}$ & $Q_{2R}$ & $Q_{em}$ & $B-L$ & Field \\
 & $\times USp(2)\times
USp(4)\times USp(2)$
& & & & & & \\
\hline\hline
$ab$ & $3 \times (4,\overline{2},1,1,1,1)$ & 1 & $-1$ & 0  & $-\frac 13,\; \frac 23,\;-1,\; 0$ & $\frac 13,\;-1$ & $Q_L, L_L$\\
$ac$ & $3 \times (\overline{4},1,2,1,1,1)$ & $-1$ & 0 & $1$   & $\frac 13,\; -\frac 23,\;1,\; 0$ & $-\frac 13,\;1$ & $Q_R, L_R$\\
$bc'$ & $2 \times(1,2,2,1,1,1)$ & 0 & $1$ & $1$   & $1,\;0,\;0,\;-1$ & 0 & $H$\\
$a1$ & $1\times (4,1,1,2,1,1)$ & $1$ & 0 & 0 & $\frac 16,\;-\frac 12$ & $\frac 13,\;-1$ & \\
$a2$ & $1\times (\overline{4},1,1,1,4,1)$ & $-1$ & 0 & 0   & $-\frac 16,\;\frac 12$ & $-\frac 13,\;1$ & \\
$a3$ & $1\times (4,1,1,1,1,\overline{2})$ & $1$ & 0 & 0   & $\frac 16,\;-\frac 12$ & $\frac 13,\;-1$ & \\
$b1$ & $2\times (1,\overline{2},1,2,1,1)$ & $0$ & -1 & 0   & $\mp\frac 12$ & 0 & \\
$b2$ & $1\times(1,2,1,1,4,1)$ & 0 & 1 & 0   & $\pm \frac 12$ & 0 & \\
$c3$ & $4\times(1,1,2,1,1,2)$ & 0 & 0 & 1   & $\pm \frac 12$ & 0 & \\
$a_{\Yasymm}$ & $4\times(6,1,1,1,1,1)$ & 2 & 0 & 0   & $-\frac 13, 1$ & $-\frac 23,2$ & \\
$b_{\Ysymm}$ & $1\times(1,3,1,1,1,1)$ & 0 & $2$ & 0   & $0,\pm 1$ & 0 & \\
$b_{\overline{\Yasymm}}$ & $1\times(1,\overline{1},1,1,1,1)$ & 0 & -2 & 0   & 0 & 0 & \\
$c_{\overline{\Ysymm}}$ & $3\times(1,1,\overline{3},1,1,1)$ & 0 & 0 & -2   & $0,\pm 1$ & 0 & \\
$c_{\Yasymm}$ & $3\times(1,1,1,1,1,1)$ & 0 & 0 & 2   & 0 & 0 & \\
\hline
\end{tabular}
\end{center}
\end{table}

\begin{table}
[htb] \footnotesize
\renewcommand{\arraystretch}{1.0}
\caption{The chiral spectrum in the open string sector of model
I-Z-5} \label{spectrum I-Z-5}
\begin{center}
\begin{tabular}{|c||c||c|c|c||c|c|c|}\hline
I-Z-5 & $SU(4)\times SU(2)_L\times SU(2)_R \times USp(2)^3$
& $Q_4$ & $Q_{2L}$ & $Q_{2R}$ & $Q_{em}$ & $B-L$ & Field \\
\hline\hline
$ab$ & $3 \times (4,\overline{2},1,1,1,1)$ & 1 & $-1$ & 0  & $-\frac 13,\; \frac 23,\;-1,\; 0$ & $\frac 13,\;-1$ & $Q_L, L_L$\\
$ac$ & $3 \times (\overline{4},1,2,1,1,1)$ & $-1$ & 0 & $1$   & $\frac 13,\; -\frac 23,\;1,\; 0$ & $-\frac 13,\;1$ & $Q_R, L_R$\\
$bc'$ & $3\times(1,\overline{2},\overline{2},1,1,1)$ & 0 & $-1$ & $-1$   & $-1,\;0,\;0,\;1$ & 0 & $H$\\
$a1$ & $1\times (4,1,1,\overline{2},1,1)$ & $1$ & 0 & 0 & $\frac 16,\;-\frac 12$ & $\frac 13,\;-1$ & \\
$a2$ & $1\times (\overline{4},1,1,1,2,1)$ & $-1$ & 0 & 0   & $-\frac 16,\;\frac 12$ & $-\frac 13,\;1$ & \\
$b2$ & $1\times(1,2,1,1,\overline{2},1)$ & 0 & 1 & 0   & $\pm \frac 12$ & 0 & \\
$c1$ & $2\times(1,1,\overline{2},2,1,1)$ & 0 & 0 & $-1$   & $\pm \frac 12$ & 0 & \\
$c3$ & $3\times(1,1,2,1,1,\overline{2})$ & 0 & 0 & 1   & $\pm \frac 12$ & 0 & \\
$b_{\Ysymm}$ & $2\times(1,3,1,1,1,1)$ & 0 & $2$ & 0   & $0,\pm 1$ & 0 & \\
$b_{\overline{\Yasymm}}$ & $2\times(1,\overline{1},1,1,1,1)$ & 0 & -2 & 0   & 0 & 0 & \\
$c_{\overline{\Ysymm}}$ & $1\times(1,1,\overline{3},1,1,1)$ & 0 & 0 & -2   & $0,\pm 1$ & 0 & \\
$c_{\Yasymm}$ & $1\times(1,1,1,1,1,1)$ & 0 & 0 & $2$   & 0 & 0 & \\
\hline
\end{tabular}
\end{center}
\end{table}

\begin{table}
[htb] \footnotesize
\renewcommand{\arraystretch}{1.0}
\caption{The chiral spectrum in the open string sector of model
I-Z-10} \label{spectrum I-Z-10}
\begin{center}
\begin{tabular}{|c||c||c|c|c||c|c|c|}\hline
I-Z-10 & $SU(4)\times SU(2)_L\times SU(2)_R \times USp(2)^4$
& $Q_4$ & $Q_{2L}$ & $Q_{2R}$ & $Q_{em}$ & $B-L$ & Field \\
\hline\hline
$ab$ & $3 \times (4,\overline{2},1,1,1,1,1)$ & 1 & -1 & 0  & $-\frac 13,\; \frac 23,\;-1,\; 0$ & $\frac 13,\;-1$ & $Q_L, L_L$\\
$ac$ & $3\times (\overline{4},1,2,1,1,1,1)$ & -1 & 0 & $1$   & $\frac 13,\; -\frac 23,\;1,\; 0$ & $-\frac 13,\;1$ & $Q_R, L_R$\\
$a1$ & $1\times (4,1,1,2,1,1,1)$ & $1$ & 0 & 0 & $\frac 16,\;-\frac 12$ & $\frac 13,\;-1$ & \\
$a2$ & $1\times (\overline{4},1,1,1,2,1,1)$ & -1 & 0 & 0   & $-\frac 16,\;\frac 12$ & $-\frac 13,\;1$ & \\
$b2$ & $1\times(1,2,1,1,2,1,1)$ & 0 & 1 & 0   & $\pm \frac 12$ & 0 & \\
$b4$ & $3\times(1,\overline{2},1,1,1,1,2)$ & 0 & -1 & 0   & $\mp \frac 12$ & 0 & \\
$c1$ & $1\times(1,1,\overline{2},2,1,1,1)$ & 0 & 0 & -1   & $\pm \frac 12$ & 0 & \\
$c3$ & $3\times(1,1,2,1,1,2,1)$ & 0 & 0 & 1   & $\pm \frac 12$ & 0 & \\
$b_{\Ysymm}$ & $2\times(1,3,1,1,1,1,1)$ & 0 & 2 & 0   & $0,\pm 1$ & 0 & \\
$b_{\overline{\Yasymm}}$ & $2\times(1,\overline{1},1,1,1,1,1)$ & 0 & -2 & 0   & 0 & 0 & \\
$c_{\overline{\Ysymm}}$ & $2\times(1,1,\overline{3},1,1,1,1)$ & 0 & 0 & -2   & $0,\pm 1$ & 0 & \\
$c_{\Yasymm}$ & $2\times(1,1,1,1,1,1,1)$ & 0 & 0 & 2   & 0 & 0 & \\
\hline
\end{tabular}
\end{center}
\end{table}

\begin{table}
[htb] \footnotesize
\renewcommand{\arraystretch}{1.0}
\caption{The composite particle spectrum of Model I-Z-10, which is
formed due to the strong forces from hidden sector.}
\label{Composite Particles I-Z-10}
\begin{center}
\begin{tabular}{|c|c||c|c|}\hline
\multicolumn{2}{|c||}{Model I-Z-10} &
\multicolumn{2}{c|}{$SU(4)\times SU(2)_L\times SU(2)_R \times
USp(2)^4$} \\
\hline Confining Force & Intersection & Exotic Particle
Spectrum & Confined Particle Spectrum \\
\hline\hline
$USp(2)_1$ &$a1$ & $1\times (4,1,1,2,1,1,1)$ & $1\times (4^2,1,1,1,1,1,1)$, $1\times(4,1,\overline{2},1,1,1,1)$\\
           &$c1$ & $1\times(1,1,\overline{2},2,1,1,1)$ & $1\times(1,1,\overline{2}^2,1,1,1,1)$\\
\hline
$USp(2)_2$ &$a2$ & $1\times (\overline{4},1,1,1,2,1,1)$ & $1\times (\overline{4}^2,1,1,1,1,1,1)$, $1\times(\overline{4},2,1,1,1,1,1)$\\
           &$b2$ & $1\times(1,2,1,1,2,1,1)$ &  $1\times(1,2^2,1,1,1,1,1)$\\
\hline
$USp(2)_3$ &$c3$ & $3\times(1,1,2,1,1,2,1)$ &  $6\times(1,1,2^2,1,1,1,1)$\\
\hline
$USp(2)_4$ &$b4$ & $3\times(1,\overline{2},1,1,1,1,2)$ & $6\times(1,\overline{2}^2,1,1,1,1,1)$\\
\hline
\end{tabular}
\end{center}
\end{table}

Hidden sector gauge groups with negative beta functions have a
potential to be confining.  In these cases, the gaugino condensations
generate the non-perturbative effective superpotential. The
minimization of this supergravity potential determines the ground
state which can stabilize the dilaton and complex structure
toroidal moduli, and in some cases breaks supersymmetry.  For the
models with two confining $USp(N)$ gauge groups, a general analysis
of the nonperturbative superpotential with tree-level gauge
couplings shows \cite{CLW} that there can be extrema with the
dilaton and complex structure moduli stabilized, however, the
extrema are saddle points in general and they do not break
supersymmetry in general. On the other hand, for  the models with
three or four confining $USp(N)$ gauge groups,  the
non-perturbative superpotientil in general allows for the
stabilization of moduli and breaking of supersymmetry at the
stable extremum. [For an explicit analysis of three  confining USp
gauge group factors see Ref. \cite{CLW}.] In our case, six models
(I-NZ-1a, I-Z-2, I-Z-3, I-Z-7, I-Z-8 and I-Z-9)
 have two $USp(N)$ gauge groups with negative beta
functions, three models (I-Z-1, I-Z-4 and I-Z-5) have three
confining $USp(N)$ gauge groups, and two models (I-Z-6 and I-Z-10)
have four confining $USp(N)$ gauge groups. Therefore, for the
latter five models, due to gaugino condensations there  may be
stable extrema with the stabilized  moduli and broken supersymmetry.
These five models are also very interesting from other
phenomenological points of view. Note also that similar to the
case studied  in Ref.~\cite{CLW}, at the extremum the cosmological
constant is likely to be negative and close to the string scale,
and thus in these models the gaugino condensation is not likely
to address the cosmological constant problem.

We should emphasize that all the models possess exotic particles
charged under the  hidden gauge group factors. There is a
possibility that the strong coupling dynamics of the hidden sector
at some intermediate scale would provide a mechanism for all of
these particles to form bound states or composite particles
(compatible with anomaly cancellation constraints); these
particles would in turn be charged only under the SM gauge
symmetry~\cite{CLS1}, which is similar to the quark condensation
in QCD. Generally speaking, $USp$ groups have two kinds of neutral
bound states. The first one is the pseudo inner product of two
fundamental representations, which is generated by reducing the
rank 2 anti-symmetric representation and is generic for $USp$
groups. This is somehow the reminiscent of a meson, which is the
inner product of one pair of $SU(3)$ fundamental and
anti-fundamental representations in QCD. The second one is rank
$2N$ anti-symmetric representation for $USp(2N)$ group with $N\geq
2$, which is also a singlet under $USp(2N)$ transformation and is
similar to a  baryon, a rank 3 anti-symmetric representation in
QCD. A definite example containing such a singlet is model I-Z-2
whose groups in the hidden sector are $USp(4)\times USp(4)$. For
$N=1$, this singlet is identified with the first one. In order to
explicitly show how to form the composite states, one concrete
example from Model I-Z-10 is given out, with the confined particle
spectra tabulated in Table \ref{Composite Particles I-Z-10}. Model
I-Z-10 has four confining gauge groups. Thereinto, both $USp(2)_1$
and $USp(2)_2$ have two charged intersections. So for them,
besides self-confinement, the mixed-confinement between different
intersections is also possible, which yields the chiral
supermultiplets $(4,1,\overline{2},1,1,1,1)$ and
$(\overline{4},2,1,1,1,1,1)$. As for $USp(2)_3$ and $USp(2)_4$,
they have only one charged intersection. Thus, there is no
mixed-confinement, and self-confinement leads to 6 tensor
representations for both ones. In addition, it is not hard to
check from the spectra that no new anomaly is introduced to the
remaining gauge symmetry, $i.e.$, this model is still
anomaly-free. This set of analysis works as well for the other
models except Model I-NZ-1(a-c) and Model I-Z-8 where a
non-asymptotical free gauge symmetry appears in the hidden sector,
and the states charged under this symmetry have no chance to be
confined. By the way, the anomaly-cancellation for the confined
particle spectra is not automatically guaranteed. Sometimes one
additional field associated with composite states is also required
to satisfy t' Hooft anomaly matching condition. Here, we only
consider one relatively simple example in order to avoid the
unnecessary complications.

\section{Some Other Potentially Interesting  Setups}

In addition to the models that we have discussed, there  are some
other potentially interesting constructions that could lead to the
SM constructions. For example, if we assume
\begin{eqnarray}
I_{ac}=-(3+h)~,~ I_{ac'}=h~,~\,
\end{eqnarray}
with $h$ a positive integer, we can have massless vector-like
Higgs fields which can break the Pati-Salam gauge symmetry down to
the SM gauge symmetry or
 break the $U(1)_{B-L}\times U(1)_{I_{3R}}$ down to $U(1)_Y$. But,
due to the large wrapping numbers required by the increased ABSes
of $I_{ac}$ and $I_{ac'}$, it is very difficult to find such
models. For instance, let us focus on the $h=1$ case with one
tilted two-torus. Considering that all factors of $2^kI_{ac}=-8$
and $2^kI_{ac'}=2$ are even, they should come from the 3rd tilted
two-torus, {\it i.e.}, we should have $|n_a^3l_c^3|=5$ or 3 and
$|n_c^3l_a^3|=3$ or 5. This implies that the 1st and 2nd WNPs will
contribute a unit factor to both $2^kI_{ac}=-8$ and
$2^kI_{ac'}=2$, and thus  we must have $n_a^1l_a^1=0$ and
$n_c^2l_c^2=0$. Therefore both a- and c-branes are of Z-type. If
$|n_a^3l_a^3|=15$, obviously there is no solution due to the TCC
violation. If $|n_c^3l_c^3|=15$, given $|n_b^3l_b^3|=1$ to
generate the even factors of $2^kI_{ab}=6$ and $2^kI_{ab'}=0$ and
the constraints from BTCC and CTCC, the ATCC cannot be satisfied.
If $|n_a^3l_a^3|=5$, to avoid the TCC violation, the even factors
of $2^kI_{ab}=6$ and $2^kI_{ab'}=0$ can not be generated from the
third WNPs at the same time. As for $|n_a^3l_a^3|=3$, there are
less than two asymptotically free  $USp$ groups available in order
to satisfy ATCC or BTCC for $|n_a^3n_c^3|=15$ or
$|l_a^3l_c^3|=15$, respectively. In short, it is hard to find such
models.

Another interesting possibility would correspond to constructions
where the $SU(2)_L$ and/or $SU(2)_R$ gauge symmetries come from
filler branes, {\it i.e.}  the $SU(2)_{L,R}=USp(2)_{L,R}$. In this
case the number of the SM Higgs doublet pairs may be decreased.
However, we do not want to obtain $SU(2)_{L,R}$ from the
splittings of higher rank $USp(N)$ ($N\ge 4$) branes which would
generically lead to even number of families.  In this case, one
wrapping number with its ABS larger than 2 cannot be avoided for
the $U(4)$ brane. This may make the model building very hard due
to tadpole cancellation constraints. However, further exploration
of these models is needed.

\section{Discussions and Conclusions}

In this paper we  reviewed the rules for supersymmetric model
building, and the conditions for the tadpole cancellations and
4-dimensional $N=1$ supersymmetric D6-brane configurations in the
Type IIA theory on $T^6/(\IZ_2\times \IZ_2)$ orientifold with
D6-brane intersections. Subsequently, we  highlighted the
interesting features of the three-family supersymmetric
$SU(4)_C\times SU(2)_L \times SU(2)_R$ models where all the gauge
symmetries arise from the stacks of D6-branes with  $U(n)$ gauge
symmetries. In particular, we demonstrated that the Pati-Salam
gauge symmetry can be broken down to the $SU(3)_C\times
SU(2)_L\times U(1)_{B-L} \times U(1)_{I_{3R}}$ via D6-brane
splittings, and further down to the Standard Model gauge symmetry
via the D- and F-flatness preserving Higgs mechanism where Higgs
fields arise from the massless  open string states in a specific
$N=2$ subsector of the theory. In order to  stabilize the
(complex structure) modulus and provide a possibility  to break
supersymmetry  via a ``race-track'' scenario, we required that
there be at least two  confining $USp$ groups in the hidden
sector.

In order to facilitate the systematic search, we  discussed the
T-duality and its variations that are in effect for the
intersecting D6-brane constructions on Type IIA orientifolds.
Employing these  symmetries allowed us to search  for all
inequivalent models  with above specified properties. We found no
models in the case  when  zero, two, or three  two-tori are
tilted. For the case with only  one tilted two-torus, we obtain 11
inequivalent models. Eight of them have  8 or fewer  pairs  of the
SM Higgs  doublets. Especially, one model has  only 2 pairs of the
SM Higgs doublets.  Furthermore, two models have the gauge
coupling unification for $SU(2)_L$ and $SU(2)_R$ gauge group
factors  at the string scale and the Higgs pairs for them arise
from the massless open string states in a $N=2$ subsector. The
explicit brane configurations, their intersections, the gauge
group structure (and the hidden sector beta functions) are
tabulated for all these models in the Appendix. As explicit
examples, we also present the chiral spectra in the open  string
 sector for the models I-NZ-1a, I-Z-6 and I-Z-10.  We  also
briefly comment on the other potentially interesting
configurations of intersecting D6-branes
  which could lead to
the  three family supersymmetric Standard Model.  In particular,
the setup when the origin of $SU(2)_L$ and/or  $SU(2)_R$  comes
from the $USp$ brane configurations is extremely constraining;  it
seems to be  very difficult to find the supersymmetric
three-family models of this type.

Models presented in this paper provide a promising stepping stone
toward the realistic SM models from string theory. In particular,
the symmetry breaking chain of the original Pati-Salam models (via
brane splitting and subsequent Higgs mechanism)  allows for
obtaining only the Standard Model gauge group structure at
electroweak scale. While we  made some preliminary comments on the
phenomenological features of these models constructed, there are a
number of avenues opened for further study.

One should study  the renormalization group equations for the
running of the gauge couplings,  both in the observable and hidden
sectors. We expect that due to the exotic matters and the
additional adjoint chiral  superfields, the low energy predictions
for the SM gauge couplings may not be consistent with these from
experiments.
  Note however,
that the left-right gauge coupling unification at the string scale
for some models  may provide interesting consequences for the low
energy couplings.

The issue of supersymmetry breaking and modulus stabilization via
gaugino condensation in the hidden sector (with at least two
confining $USp$ gauge group factors) should be addressed in
detail.  If supersymmetry breaking does take place in these
models, this breaking  can be mediated via gauge interactions,
thus providing a possibility to  address  the CP problem in this
framework.  In addition, the nature of the soft supersymmetry
breaking parameters and their model dependence, deserve further
detailed study.   Another important role of the strong dynamics in
the hidden sector could play, is to bind fractionally charged
exotic matter into the composite objects  with only SM quantum
numbers~\cite{CLS1}.

An important topic is further study of the Higgs mechanism (that
preserves D- and F-flatness  condition) for the breaking of
$U(1)_{B-L}\times U(1)_{I_{3R}}$  down to $U(1)_Y$. In order to be
consistent with the see-saw mechanism to explain the tiny neutrino
masses, and with the leptogenesis to produce the baryon asymmetry,
we need the symmetry breaking scale at about $10^{15}$ GeV which
hopefully can be realized in our models (with radiative
corrections fixing this scale as one possibility).

The study of the SM fermion masses and mixings is also important.
Yukawa couplings can be calculated exactly in conformal field
theory \cite{MCIP} and they have a beautiful geometric
interpretation in terms of the angles and areas of the triangles,
specified by the location of brane intersections in the internal
space. In particular, the models with few SM Higgs pairs should
provide an important framework to address the textures of the SM
fermion mass matrices
 in  detail \cite{CLLL}.

\section*{Acknowledgments}

We would like to thank Paul  Langacker for useful discussions. The
research was supported  in part by the National
 Science Foundation under Grant No.~INT02-03585 (MC)
 and PHY-0070928 (T. Li),  DOE-EY-76-02-3071 (MC, T. Liu)
and  Fay R. and Eugene L. Langberg Chair  (MC).

\newpage

\begin{center}
\Large{\bf Appendix: Tables for Supersymmetric Pati-Salam Models}
\end{center}

In Appendix  we tabulate all 11 inequivalent models, found by the
systematic search. Thereinto, we present only its equivalence class,
specified by T-dualities. In the first
column of each table, $a$, $b$ and $c$ denote the
$U(4)$, $U(2)_L$, and $U(2)_R$ stacks of branes,
 respectively. 1, 2, 3, and 4 represent the filler branes
along respective $\Omega R$, $\Omega R\omega$, $\Omega
R\theta\omega$ and $\Omega R\theta$ orientifold planes, resulting
in $USp(N)$ gauge groups.
 $N$ in the second column is the number of D6-branes in each stack.
The third column shows the wrapping numbers of the various branes,
and we have specified the third set of wrapping number for the
tilted two-torus. (Recall, only one two-torus is tilted.)  The
intersection numbers between the various stacks are given in the
remaining right columns where $b'$ and $c'$ are respectively the $\Omega R$
images of $b$ and $c$.
For convenience, we also tabulate  the
relation among the  moduli parameters imposed by the supersymmetry
conditions, as well as the $\beta$ functions ($\beta^g_i$) of the gauge
groups in the
hidden sector. Again,  we required   at least two
asymptotically free $USp$ groups with negative $\beta$ functions in
the hidden sector. For example, model I-NZ-1a, which forms one
equivalent class together with I-NZ-1b and I-NZ-1c, has two
asymptotically free $USp$ groups among its three $USp$ groups in the
hidden sector. Moreover, I-, -Z- and -NZ- imply  that
 only one two-torus is tilted,  a-brane is
Z-type, and a-brane is
NZ-type, respectively.

\begin{table}
[htb] \footnotesize
\renewcommand{\arraystretch}{1.0}
\caption{D6-brane configurations and intersection numbers for the
three-family left-right symmetric model I-NZ-1a} \label{model
I-NZ-1a}
\begin{center}
\begin{tabular}{|c||c|c||c|c|c|c|c|c|c|c|c|}
\hline
    \rm{model} I-NZ-1a & \multicolumn{11}{c|}{$U(4)\times U(2)_L\times U(2)_R\times USp(2)\times USp(4) \times
    USp(2)$}\\
\hline \hline \rm{stack} & $N$ & $(n^1,l^1)\times (n^2,l^2)\times
(n^3,l^3)$ & $n_{\Ysymm}$& $n_{\Yasymm}$ & $b$ & $b'$ & $c$ & $c'$& 1 & 2 & 3 \\
\hline
    $a$&  8& $(1,-1)\times (1,1)\times (1,1)$ & 0 & 4  & 3 & 0 & -3 & 0 & 1 & -1 & 1 \\
    $b$&  4& $(0,1)\times (1,-2)\times (1,-1)$ & 1 & -1  & - & - & 0 & 2 & -2 & 1 & 0  \\
    $c$&  4& $(1,0)\times (1,4)\times (1,-1)$ & -3 & 3  & - & - & - & - & 0 & 0 & 4 \\
\hline
    1&   2& $(1,0)\times (1,0)\times (2,0)$& \multicolumn{9}{c|}{$X_A = {1 \over 2}X_B = 5X_C = {5 \over 4}X_D$}\\
    2&   4& $(1,0)\times (0,-1)\times (0,2)$ & \multicolumn{9}{c|}{$(\chi_1=5/\sqrt{2},\chi_2=1/2\sqrt{2},\chi_3=2\sqrt{2})$}\\
    3&   2& $(0,-1)\times (1,0)\times (2,0)$ & \multicolumn{9}{c|}{$\beta^g_1=-2, \beta^g_2=-3, \beta^g_3=0$}\\
\hline
\end{tabular}
\end{center}
\end{table}

\begin{table}
[htb] \footnotesize
\renewcommand{\arraystretch}{1.0}
\caption{D6-brane configurations and intersection numbers for the
three-family left-right symmetric model I-NZ-1b} \label{model
I-NZ-1b}
\begin{center}
\begin{tabular}{|c||c|c||c|c|c|c|c|c|c|c|c|}
\hline
    \rm{model} I-NZ-1b & \multicolumn{11}{c|}{$U(4)\times U(2)_L\times U(4)_R\times USp(2)\times USp(2) \times
    USp(2)$}\\
\hline \hline \rm{stack} & $N$ & $(n^1,l^1)\times (n^2,l^2)\times
(n^3,l^3)$ & $n_{\Ysymm}$& $n_{\Yasymm}$ & $b$ & $b'$ & $c$ & $c'$& 2 & 3 & 4 \\
\hline
    $a$&  8& $(1,-1)\times (1,1)\times (1,1)$ & 0 & 4  & 3 & 0 & -3 & 0 & -1 & 1 & 1 \\
    $b$&  4& $(2,1)\times (1,0)\times (1,-1)$ & 1 & -1  & - & - & 0 & 2 & 1 & 0 & -2  \\
    $c$&  4& $(4,-1)\times (0,1)\times (1,-1)$ & -3 & 3  & - & - & - & - & 0 & 4 & 0 \\
\hline
    2&   4& $(1,0)\times (0,-1)\times (0,2)$ & \multicolumn{9}{c|}{$X_A = {2 \over 5}X_B = 4X_C = {4 \over 5}X_D$}\\
    3&   2& $(0,-1)\times (1,0)\times (2,0)$ & \multicolumn{9}{c|}{$(\chi_1=2\sqrt{2},\chi_2= \sqrt{2}/5,\chi_3=2\sqrt{2})$}\\
    4&   2& $(0,-1)\times (0,1)\times (2,0)$ & \multicolumn{9}{c|}{$\beta^g_2=-3, \beta^g_3=0, \beta^g_4=-2$}\\
\hline
\end{tabular}
\end{center}
\end{table}

\begin{table}
[htb] \footnotesize
\renewcommand{\arraystretch}{1.0}
\caption{D6-brane configurations and intersection numbers for the
three-family left-right symmetric model I-NZ-1c} \label{model
I-NZ-1c}
\begin{center}
\begin{tabular}{|c||c|c||c|c|c|c|c|c|c|c|c|}
\hline
    \rm{model} I-NZ-1c & \multicolumn{11}{c|}{$U(4)\times U(2)_L\times U(2)_R\times USp(4)\times USp(2) \times
    USp(2)$}\\
\hline \hline \rm{stack} & $N$ & $(n^1,l^1)\times (n^2,l^2)\times
(n^3,l^3)$ & $n_{\Ysymm}$& $n_{\Yasymm}$ & $b$ & $b'$ & $c$ & $c'$& 1 & 2 & 4 \\
\hline
    $a$&  8& $(-1,-1)\times (1,1)\times (1,1)$ & 0 & -4  & 3 & 0 & -3 & 0 & 1 & -1 & -1 \\
    $b$&  4& $(1,0)\times (4,1)\times (1,-1)$ & 3 & -3  & - & - & 0 & -2 & 0 & 0 & -4  \\
    $c$&  4& $(0,1)\times (2,-1)\times (1,-1)$ & -1 & 1  & - & - & - & - & -1 & 2 & 0 \\
\hline
    1&   4& $(1,0)\times (1,0)\times (2,0)$& \multicolumn{9}{c|}{$X_A = 2X_B = {5\over 2}X_C = 10X_D$}\\
    2&   2& $(1,0)\times (0,-1)\times (0,2)$ & \multicolumn{9}{c|}{$(\chi_1=5/ \sqrt{2},\chi_2=2 \sqrt{2} ,\chi_3=\sqrt{2})$}\\
    4&   2& $(0,-1)\times (0,1)\times (2,0)$ & \multicolumn{9}{c|}{$\beta^g_1=-3, \beta^g_2=-2, \beta^g_4=0$}\\
\hline
\end{tabular}
\end{center}
\end{table}

\begin{table}
[htb] \footnotesize
\renewcommand{\arraystretch}{1.0}
\caption{D6-brane configurations and intersection numbers for the
three-family left-right symmetric model I-Z-1} \label{model I-Z-1}
\begin{center}
\begin{tabular}{|c||c|c||c|c|c|c|c|c|c|c|c|}
\hline
    \rm{model} I-Z-1 & \multicolumn{11}{c|}{$U(4)\times U(2)_L\times U(2)_R\times USp(2)\times USp(2) \times
    USp(2)$}\\
\hline \hline \rm{stack} & $N$ & $(n^1,l^1)\times (n^2,l^2)\times
(n^3,l^3)$ & $n_{\Ysymm}$& $n_{\Yasymm}$ & $b$ & $b'$ & $c$ & $c'$& 1 & 2 & 3 \\
\hline
    $a$&  8& $(0,-1)\times (1,1)\times (1,1)$ & 0 & 0  & 3 & 0 & -3 & 0 & 1 & -1 & 0 \\
    $b$&  4& $(1,0)\times (4,1)\times (1,-1)$ & 3 & -3  & - & - & 0 & -7 & 0 & 0 & 1  \\
    $c$&  4& $(-1,1)\times (1,-2)\times (1,-1)$ & 2 & -6  & - & - & - & - & -2 & 1 & 2 \\
\hline
    1&   2& $(1,0)\times (1,0)\times (2,0)$& \multicolumn{9}{c|}{$X_A = X_B = {9 \over 4}X_C = 9X_D$}\\
    2&   2& $(1,0)\times (0,-1)\times (0,2)$ & \multicolumn{9}{c|}{$\beta^g_1=-2, \beta^g_2=-3, \beta^g_3=-3$}\\
    3&   2& $(0,-1)\times (1,0)\times (2,0)$ & \multicolumn{9}{c|}{}\\
\hline
\end{tabular}
\end{center}
\end{table}

\begin{table}
[htb] \footnotesize
\renewcommand{\arraystretch}{1.0}
\caption{D6-brane configurations and intersection numbers for the
three-family left-right symmetric model I-Z-2} \label{model I-Z-2}
\begin{center}
\begin{tabular}{|c||c|c||c|c|c|c|c|c|c|c|}
\hline
    \rm{model} I-Z-2 & \multicolumn{10}{c|}{$U(4)\times U(2)_L\times U(2)_R\times USp(4)\times
    USp(4)$}\\
\hline \hline \rm{stack} & $N$ & $(n^1,l^1)\times (n^2,l^2)\times
(n^3,l^3)$ & $n_{\Ysymm}$& $n_{\Yasymm}$ & $b$ & $b'$ & $c$ & $c'$& 1 & 2  \\
\hline
    $a$&  8& $(0,-1)\times (1,1)\times (1,1)$ & 0 & 0  & 3 & 0 & -3 & 0 & 1 & -1  \\
    $b$&  4& $(3,1)\times (1,0)\times (1,-1)$ & 2 & -2  & - & - & 0 & -4 & 0 & 1   \\
    $c$&  4& $(-1,1)\times (1,-2)\times (1,-1)$ & 2 & 6  & - & - & - & - & -2 & 1  \\
\hline
    1&   4& $(1,0)\times (1,0)\times (2,0)$& \multicolumn{8}{c|}{$X_A = X_B = 3X_C = 3X_D$}\\
    2&   4& $(1,0)\times (0,-1)\times (0,2)$ & \multicolumn{8}{c|}{$\beta^g_1=-2, \beta^g_2=-2$}\\
\hline
\end{tabular}
\end{center}
\end{table}

\begin{table}
[htb] \footnotesize
\renewcommand{\arraystretch}{1.0}
\caption{D6-brane configurations and intersection numbers for the
three-family left-right symmetric model I-Z-3} \label{model I-Z-3}
\begin{center}
\begin{tabular}{|c||c|c||c|c|c|c|c|c|c|c|}
\hline
    \rm{model} I-Z-3 & \multicolumn{10}{c|}{$U(4)\times U(2)_L\times U(2)_R\times USp(2)\times USp(4)$}\\
\hline \hline \rm{stack} & $N$ & $(n^1,l^1)\times (n^2,l^2)\times
(n^3,l^3)$ & $n_{\Ysymm}$& $n_{\Yasymm}$ & $b$ & $b'$ & $c$ & $c'$& 2 & 4 \\
\hline
    $a$&  8& $(0,-1)\times (1,1)\times (1,1)$ & 0 & 0  & 3 & 0 & -3 & 0 & 0 & 0  \\
    $b$&  4& $(3,1)\times (1,0)\times (1,-1)$ & 2 & -2  & - & - & 0 & 4 & 0 & -3  \\
    $c$&  4& $(1,0)\times (1,4) \times (1,-1)$ & -3 & 3  & - & - & - & - & 4 & -1 \\
\hline
    3&   2& $(0,-1)\times (1,0)\times (0,2)$& \multicolumn{8}{c|}{$X_A = X_B = 12X_C = 3X_D$}\\
    4&   4& $(0,-1)\times (0,1)\times (2,0)$ & \multicolumn{8}{c|}{$\beta^g_3=-2,
    \beta^g_4=-2$}\\
\hline
\end{tabular}
\end{center}
\end{table}

\begin{table}
[htb] \footnotesize
\renewcommand{\arraystretch}{1.0}
\caption{D6-brane configurations and intersection numbers for the
three-family left-right symmetric model I-Z-4} \label{model I-Z-4}
\begin{center}
\begin{tabular}{|c||c|c||c|c|c|c|c|c|c|c|c|}
\hline
    \rm{model} I-Z-4 & \multicolumn{11}{c|}{$U(4)\times U(2)_L\times U(2)_R\times USp(2)\times USp(2) \times
    USp(2)$}\\
\hline \hline \rm{stack} & $N$ & $(n^1,l^1)\times (n^2,l^2)\times
(n^3,l^3)$ & $n_{\Ysymm}$& $n_{\Yasymm}$ & $b$ & $b'$ & $c$ & $c'$& 1 & 2 & 3 \\
\hline
    $a$&  8& $(0,-1)\times (1,1)\times (1,1)$ & 0 & 0  & 2 & 1 & -3 & 0 & 1 & -1 & 0 \\
    $b$&  4& $(1,1)\times (1,0)\times (3,-1)$ & -2 & 2  & - & - & 5 & 2 & 0 & 3 & 0  \\
    $c$&  4& $(3,-2)\times (0,1)\times (1,-1)$ & -1 & 1  & - & - & - & - & -2 & 0 & 3 \\
\hline
    1&   2& $(1,0)\times (1,0)\times (2,0)$& \multicolumn{9}{c|}{$X_A = X_B = {3 \over 2}X_C = {1\over 3}X_D$}\\
    2&   2& $(1,0)\times (0,-1)\times (0,2)$ & \multicolumn{9}{c|}{$\beta^g_1=-2, \beta^g_2=-1, \beta^g_3=-3$}\\
    3&   2& $(0,-1)\times (1,0)\times (2,0)$ & \multicolumn{9}{c|}{}\\
\hline
\end{tabular}
\end{center}
\end{table}

\begin{table}
[htb] \footnotesize
\renewcommand{\arraystretch}{1.0}
\caption{D6-brane configurations and intersection numbers for the
three-family left-right symmetric model I-Z-5} \label{model I-Z-5}
\begin{center}
\begin{tabular}{|c||c|c||c|c|c|c|c|c|c|c|c|}
\hline
    \rm{model} I-Z-5 & \multicolumn{11}{c|}{$U(4)\times U(2)_L\times U(2)_R\times USp(2)\times USp(2) \times
    USp(2)$}\\
\hline \hline \rm{stack} & $N$ & $(n^1,l^1)\times (n^2,l^2)\times
(n^3,l^3)$ & $n_{\Ysymm}$& $n_{\Yasymm}$ & $b$ & $b'$ & $c$ & $c'$& 1 & 2 & 3 \\
\hline
    $a$&  8& $(0,-1)\times (1,1)\times (1,1)$ & 0 & 0  & 3 & 0 & -3 & 0 & 1 & -1 & 0 \\
    $b$&  4& $(3,1)\times (1,0)\times (1,-1)$ & 2 & -2  & - & - & 0 & -3 & 0 & 1 & 0  \\
    $c$&  4& $(3,-2)\times (0,1)\times (1,-1)$ & -1 & 1  & - & - & - & - & -2 & 0 & 3 \\
\hline
    1&   2& $(1,0)\times (1,0)\times (2,0)$& \multicolumn{9}{c|}{$X_A = X_B = {3 \over 2}X_C = 3X_D$}\\
    2&   2& $(1,0)\times (0,-1)\times (0,2)$ & \multicolumn{9}{c|}{$\beta^g_1=-2, \beta^g_2=-3, \beta^g_3=-3$}\\
    3&   2& $(0,-1)\times (1,0)\times (2,0)$ & \multicolumn{9}{c|}{}\\
\hline
\end{tabular}
\end{center}
\end{table}

\begin{table}
[htb] \footnotesize
\renewcommand{\arraystretch}{1.0}
\caption{D6-brane configurations and intersection numbers for the
three-family left-right symmetric model I-Z-6} \label{model I-Z-6}
\begin{center}
\begin{tabular}{|c||c|c||c|c|c|c|c|c|c|c|c|c|}
\hline
    \rm{model} I-Z-6 & \multicolumn{12}{c|}{$U(4)\times U(2)_L\times U(2)_R\times USp(2)\times USp(2)\times USp(2) \times
    USp(2)$}\\
\hline \hline \rm{stack} & $N$ & $(n^1,l^1)\times (n^2,l^2)\times
(n^3,l^3)$ & $n_{\Ysymm}$& $n_{\Yasymm}$ & $b$ & $b'$ & $c$ & $c'$& 1 & 2 & 3 & 4 \\
\hline
    $a$&  8& $(0,-1)\times (1,1)\times (1,1)$ & 0 & 0  & 2 & 1 & -3 & 0 & 1 & -1 & 0 & 0\\
    $b$&  4& $(1,1)\times (1,0)\times (3,-1)$ & -2 & 2  & - & - & 4 & 4 & 0 & 3 & 0 & -1 \\
    $c$&  4& $(3,-1)\times (0,1)\times (1,-1)$ & -2 & 2  & - & - & - & - & -1 & 0 & 3 & 0\\
\hline
    1&   2& $(1,0)\times (1,0)\times (2,0)$& \multicolumn{10}{c|}{$X_A = X_B = 3X_C = {1\over 3}X_D$}\\
    2&   2& $(1,0)\times (0,-1)\times (0,2)$ & \multicolumn{10}{c|}{$\beta^g_1=-3, \beta^g_2=-1, \beta^g_3=-3,\beta^g_4=-5$}\\
    3&   2& $(0,-1)\times (1,0)\times (0,2)$& \multicolumn{10}{c|}{}\\
    4&   2& $(0,-1)\times (0,1)\times (2,0)$ & \multicolumn{10}{c|}{}\\
\hline
\end{tabular}
\end{center}
\end{table}

\begin{table}
[htb] \footnotesize
\renewcommand{\arraystretch}{1.0}
\caption{D6-brane configurations and intersection numbers for the
three-family left-right symmetric model I-Z-7} \label{model I-Z-7}
\begin{center}
\begin{tabular}{|c||c|c||c|c|c|c|c|c|c|c|}
\hline
    \rm{model} I-Z-7 & \multicolumn{10}{c|}{$U(4)\times U(2)_L\times U(2)_R\times USp(2)\times USp(4)$}\\
\hline \hline \rm{stack} & $N$ & $(n^1,l^1)\times (n^2,l^2)\times
(n^3,l^3)$ & $n_{\Ysymm}$& $n_{\Yasymm}$ & $b$ & $b'$ & $c$ & $c'$ & 3 & 4 \\
\hline
    $a$&  8& $(0,-1)\times (1,1)\times (1,1)$ & 0 & 0  & 2 & 1 & -3 & 0  & 0 & 0\\
    $b$&  4& $(1,1)\times (1,0)\times (3,-1)$ & -2 & 2  & - & - & 4 & 8  & 0 & -1 \\
    $c$&  4& $(1,0)\times (1,4)\times (1,-1)$ & -3 & 3  & - & - & - & -  & 4 & -1\\
\hline
    3&   2& $(0,-1)\times (1,0)\times (0,2)$& \multicolumn{8}{c|}{$X_A = X_B = {4 \over 3}X_C = {1\over 3}X_D$}\\
    4&   4& $(0,-1)\times (0,1)\times (2,0)$ & \multicolumn{8}{c|}{$ \beta^g_3=-2,\beta^g_4=-4$}\\
\hline
\end{tabular}
\end{center}
\end{table}

\begin{table}
[htb] \footnotesize
\renewcommand{\arraystretch}{1.0}
\caption{D6-brane configurations and intersection numbers for the
three-family left-right symmetric model I-Z-8} \label{model I-Z-8}
\begin{center}
\begin{tabular}{|c||c|c||c|c|c|c|c|c|c|c|c|}
\hline
    \rm{model} I-Z-8 & \multicolumn{11}{c|}{$U(4)\times U(2)_L\times U(2)_R\times USp(2)\times USp(2) \times
    USp(2)$}\\
\hline \hline \rm{stack} & $N$ & $(n^1,l^1)\times (n^2,l^2)\times
(n^3,l^3)$ & $n_{\Ysymm}$& $n_{\Yasymm}$ & $b$ & $b'$ & $c$ & $c'$& 1 & 2 & 4 \\
\hline
    $a$&  8& $(0,-1)\times (1,1)\times (1,1)$ & 0 & 0  & 2 & 1 & -3 & 0 & 1 & -1 & 0\\
    $b$&  4& $(1,2)\times (1,0)\times (3,-1)$ & -5 & 5  & - & - & 7 & 10 & 0 & 6 & -1 \\
    $c$&  4& $(3,-1)\times (0,1)\times (1,-1)$ & -2 & 2  & - & - & - & - & -1 & 0 & 0\\
\hline
    1&   2& $(1,0)\times (1,0)\times (2,0)$& \multicolumn{9}{c|}{$X_A = X_B = 3X_C = {1\over 6}X_D$}\\
    2&   2& $(1,0)\times (0,-1)\times (0,2)$ & \multicolumn{9}{c|}{$\beta^g_1=-3, \beta^g_2=2,\beta^g_4=-5$}\\
    4&   2& $(0,-1)\times (0,1)\times (2,0)$ & \multicolumn{9}{c|}{}\\
\hline
\end{tabular}
\end{center}
\end{table}

\begin{table}
[htb] \footnotesize
\renewcommand{\arraystretch}{1.0}
\caption{D6-brane configurations and intersection numbers for the
three-family left-right symmetric model I-Z-9} \label{model I-Z-9}
\begin{center}
\begin{tabular}{|c||c|c||c|c|c|c|c|c|c|c|}
\hline
    \rm{model} I-Z-9 & \multicolumn{10}{c|}{$U(4)\times U(2)_L\times U(2)_R\times USp(2)\times USp(2)$}\\
\hline \hline \rm{stack} & $N$ & $(n^1,l^1)\times (n^2,l^2)\times
(n^3,l^3)$ & $n_{\Ysymm}$& $n_{\Yasymm}$ & $b$ & $b'$ & $c$ & $c'$& 1 & 2  \\
\hline
    $a$&  8& $(0,-1)\times (1,1)\times (1,1)$ & 0 & 0  & 3 & 0 & -3 & 0 & 1 & -1  \\
    $b$&  4& $(3,2)\times (1,0)\times (1,-1)$ & -1 & 1  & - & - & 0 & 0 & 0 & 2   \\
    $c$&  4& $(3,-2)\times (0,1)\times (1,-1)$ & 1 & -1  & - & - & - & - & -2 & 0  \\
\hline
    1&   2& $(1,0)\times (1,0)\times (2,0)$& \multicolumn{8}{c|}{$X_A = X_B = {3 \over 2}X_C = {3\over 2}X_D$}\\
    2&   2& $(1,0)\times (0,-1)\times (0,2)$ & \multicolumn{8}{c|}{$\beta^g_1=-2, \beta^g_2=-2,$}\\
\hline
\end{tabular}
\end{center}
\end{table}

\begin{table}
[htb] \footnotesize
\renewcommand{\arraystretch}{1.0}
\caption{D6-brane configurations and intersection numbers for the
three-family left-right symmetric model I-Z-10} \label{model
I-Z-10}
\begin{center}
\begin{tabular}{|c||c|c||c|c|c|c|c|c|c|c|c|c|}
\hline
    \rm{model} I-Z-10 & \multicolumn{12}{c|}{$U(4)\times U(2)_L\times U(2)_R\times USp(2)\times USp(2)\times USp(2) \times
    USp(2)$}\\
\hline \hline \rm{stack} & $N$ & $(n^1,l^1)\times (n^2,l^2)\times
(n^3,l^3)$ & $n_{\Ysymm}$& $n_{\Yasymm}$ & $b$ & $b'$ & $c$ & $c'$& 1 & 2 & 3 & 4 \\
\hline
    $a$&  8& $(0,-1)\times (1,1)\times (1,1)$ & 0 & 0  & 3 & 0 & -3 & 0 & 1 & -1 & 0 & 0\\
    $b$&  4& $(3,1)\times (1,0)\times (1,-1)$ & 2 & -2  & - & - & 0 & 0 & 0 & 1 & 0 & -3 \\
    $c$&  4& $(3,-1)\times (0,1)\times (1,-1)$ & -2 & 2  & - & - & - & - & -1 & 0 & 3 & 0\\
\hline
    1&   2& $(1,0)\times (1,0)\times (2,0)$& \multicolumn{10}{c|}{$X_A = X_B = 3X_C = 3X_D$}\\
    2&   2& $(1,0)\times (0,-1)\times (0,2)$ & \multicolumn{10}{c|}{$\beta^g_1=-3, \beta^g_2=-3, \beta^g_3=-3,\beta^g_4=-3$}\\
    3&   2& $(0,-1)\times (1,0)\times (0,2)$& \multicolumn{10}{c|}{}\\
    4&   2& $(0,-1)\times (0,1)\times (2,0)$ & \multicolumn{10}{c|}{}\\
\hline
\end{tabular}
\end{center}
\end{table}


\begin{thebibliography}{99}

%

\bibitem{JPEW}
J.~Polchinski and E.~Witten, Nucl.\ Phys.\ B {\bf 460}, 525
(1996).

\bibitem{ABPSS}
C.~Angelantonj, M.~Bianchi, G.~Pradisi, A.~Sagnotti and
Y.~S.~Stanev, Phys.\ Lett.\ B {\bf 385}, 96 (1996).

\bibitem{berkooz}
M.~Berkooz and R.G.~Leigh, Nucl.\ Phys.\ B {\bf 483}, 187 (1997).

\bibitem{ShiuTye}
G.~Shiu and S.~H.~Tye, Phys.\ Rev.\ D {\bf 58}, 106007 (1998).

\bibitem{lpt}
J.~Lykken, E.~Poppitz and S.~P.~Trivedi, Nucl.\ Phys.\ B {\bf
543}, 105 (1999).

\bibitem{MCJW}
M.~Cveti\v c, M.~Pl\"umacher and J.~Wang, JHEP {\bf 0004}, 004 (2000);
M.~Cveti\v c, A.~M.~Uranga and J.~Wang, Nucl.\ Phys.\ B {\bf 595}, 63
(2001).

\bibitem{Ibanez}
G.~Aldazabal, A.~Font, L.~E.~Ib\'a\~nez and G.~Violero, Nucl.\ Phys.\
B {\bf 536}, 29 (1998); G.~Aldazabal, L.~E.~Ib\'a\~nez, F.~Quevedo and
A.~M.~Uranga, JHEP {\bf 0008}, 002 (2000).

\bibitem{MKRR}
M.~Klein and R.~Rabadan, JHEP {\bf 0010}, 049 (2000).


\bibitem{bdl}
M.~Berkooz, M.~R.~Douglas and R.~G.~Leigh, Nucl. Phys. B {\bf 480}
(1996) 265.

\bibitem{bachas}
C.~Bachas, hep-th/9503030.

\bibitem{urangac}
J.~F.~G.~Cascales and A.~M.~Uranga,
hep-th/0311250.

\bibitem{bgkl}
R.~Blumenhagen, L.~G\"orlich, B.~K\"ors and D.~L\"ust, JHEP {\bf
0010} (2000) 006.

\bibitem{bkl}
R.~Blumenhagen, B.~K\"ors and D.~L\"ust, JHEP {\bf 0102} (2001)
030.

\bibitem{afiru}
G.~Aldazabal, S.~Franco, L.~E.~Ib\'a\~nez, R.~Rabad\'an and
A.~M.~Uranga, JHEP {\bf 0102}, 047 (2001).

\bibitem{Uranga}
G.~Aldazabal, S.~Franco, L.~E.~Ib\'a\~nez, R.~Rabadan and
A.~M.~Uranga, J.\ Math.\ Phys.\  {\bf 42}, 3103 (2001).

\bibitem{imr}
L.~E.~Ib\'a\~nez, F.~Marchesano and R.~Rabad\'an, JHEP {\bf 0111},
002 (2001).

\bibitem{magnetised}
C.~Angelantonj, I.~Antoniadis, E.~Dudas and A.~Sagnotti, Phys.
Lett. B {\bf 489} (2000) 223.



\bibitem{bonn}
S.~F\"orste, G.~Honecker and R.~Schreyer, Nucl. Phys. B {\bf 593}
(2001) 127; JHEP {\bf 0106} (2001) 004.

\bibitem{bklo}
R.~Blumenhagen, B.~K\"ors and D.~L\"ust, T.~Ott, Nucl. Phys. {\bf
B616} (2001) 3.

\bibitem{cim}
D.~Cremades, L.~E.~Ib\'a\~nez and F.~Marchesano, Nucl.\ Phys.\ B {\bf
643}, 93 (2002).

\bibitem{CIM1}
D.~Cremades, L.~E.~Ib\'a\~nez and F.~Marchesano, JHEP 0207, 009(2002).

\bibitem{CIM2}
D.~Cremades, L.~E.~Ib\'a\~nez and F.~Marchesano, JHEP 0207, 022(2002).

\bibitem{bailin}
D.~Bailin, G.~V.~Kraniotis, and A.~Love, Phys.\ Lett.\ B {\bf
530}, 202 (2002); Phys.\ Lett.\ B {\bf 547}, 43 (2002); Phys.\
Lett.\ B {\bf 553}, 79 (2003); JHEP {\bf 0302}, 052 (2003).

\bibitem{JREDVN}
J.~R.~Ellis, P.~Kanti and D.~V.~Nanopoulos, Nucl.\ Phys.\ B {\bf
647}, 235 (2002).

\bibitem{kokorelis}  C.~Kokorelis,
JHEP {\bf 0209}, 029 (2002); JHEP {\bf 0208}, 036 (2002);
hep-th/0207234; JHEP {\bf 0211}, 027 (2002); hep-th/0210200.

\bibitem{CSU1}
M.~Cveti\v c, G.~Shiu and A.~M.~Uranga, Phys.\ Rev.\ Lett.\  {\bf
87}, 201801 (2001).

\bibitem{CSU2}
M.~Cveti\v c, G.~Shiu and A.~M.~Uranga, Nucl.\ Phys.\ B {\bf 615},
3 (2001).

\bibitem{CP} M. Cveti\v c and I. Papadimitriou,
Phys.\ Rev.\ D {\bf 67}, 126006 (2003).

\bibitem{CPS}
M. Cveti\v c, I. Papadimitriou and G. Shiu, hep-th/0212177.


\bibitem{CLS1}
 M.~Cveti\v c, P.~Langacker and G.~Shiu,
Phys.\ Rev.\ D {\bf 66}, 066004 (2002).

\bibitem{CLS2}
M.~Cveti\v c, P.~Langacker and G.~Shiu, Nucl.\ Phys.\ B {\bf 642},
139 (2002).

\bibitem{MCIP}
M.~Cveti\v c and I.~Papadimitriou, Phys.\ Rev.\ D {\bf 68}, 046001
(2003).

\bibitem{CLW}
M.~Cveti\v c, P.~Langacker and J.~Wang, Phys.\ Rev.\ D {\bf 68},
046002 (2003).
\bibitem{blumrecent}

R.~Blumenhagen, L.~G\"orlich and T.~Ott, hep-th/0211059.

\bibitem{Honecker}
G.~Honecker, hep-th/0303015.

\bibitem{RB11}
R.~Blumenhagen, JHEP {\bf0311}, 055 (2003).


\bibitem{RB12}
R.~Blumenhagen and T.~Weigand, JHEP {\bf0402}, 041 (2004).

\bibitem{Ilke}
I.~Brunner, K.~Hori, K.~Hosomichi and J.~Walcher,
hep-th/0401137.

\bibitem{LLG3}
T.~Li and T.~Liu, Phys.\ Lett.\ B {\bf 573}, 193 (2003),
hep-th/0304258.

\bibitem{Sen}
A.~Sen, hep-th/9802051.

\bibitem{Taylor}
T.~R.~Taylor, Phys.\ Lett.\ B {\bf 252}, 59 (1990).

\bibitem{RBPJS}
R.~Brustein and P.~J.~Steinhardt, Phys.\ Lett.\ B {\bf 302}, 196
(1993).

\bibitem{BDCCM}
B.~de Carlos, J.~A.~Casas and C.~Munoz, Nucl.\ Phys.\ B {\bf 399},
623 (1993).

\bibitem{CLLL}
M. Cveti\v c, P. Langacker, T. Li and T. Liu, work in progress.
\end{thebibliography}
\end{document}